\documentclass[lettersize,journal]{IEEEtran}
\usepackage{amsmath,amsfonts}
\usepackage{algorithm}
\usepackage{array}
\usepackage[caption=false,font=normalsize,labelfont=sf,textfont=sf]{subfig}
\usepackage{textcomp}
\usepackage{stfloats}
\usepackage{url}
\usepackage{verbatim}
\usepackage{graphicx}
\usepackage{cite}
\usepackage{booktabs}
\usepackage{algorithmicx}%
\usepackage{algpseudocode}%
\begin{document}

\title{Securing Satellite Communications: Real-Time Video Encryption Scheme on Satellite Payloads}

%\author{IEEE Publication Technology,~\IEEEmembership{Staff,~IEEE,}
	
%\author{Hanshuo~Qiu, 
%	Jing~Lian,
%	Xiaoyuan~Wang,
%	and Jizhao~Liu% <-this % stops a space
%	\IEEEcompsocitemizethanks{
%		\IEEEcompsocthanksitem H. Qiu and J. Liu are with the School of Information Science and Engineering, Lanzhou University, No.222, TianShui Road(south), Lanzhou, 730000, Gansu, China.
%		\IEEEcompsocthanksitem J. Lian is with the School of Electronics and Information Engineering, Lanzhou Jiaotong University, Lanzhou, 730070, Gansu, China.
%		\IEEEcompsocthanksitem X. Wang is with the School of Electronics and Information, Hangzhou Dianzi University, Hangzhou 310018, China.
%	}% <-this % stops a space
%}
\author{Hanshuo~Qiu, 
	Jing~Lian, 
	Xiaoyuan~Wang, 
	and Jizhao~Liu\IEEEauthorrefmark{1}% <-this % stops a space
	\IEEEcompsocitemizethanks{
		\IEEEcompsocthanksitem H. Qiu and J. Liu are with the School of Information Science and Engineering, Lanzhou University, No.222, TianShui Road(south), Lanzhou, 730000, Gansu, China.
		\IEEEcompsocthanksitem J. Lian is with the School of Electronics and Information Engineering, Lanzhou Jiaotong University, Lanzhou, 730070, Gansu, China.
		\IEEEcompsocthanksitem X. Wang is with the School of Electronics and Information, Hangzhou Dianzi University, Hangzhou 310018, China.
		\IEEEcompsocthanksitem \IEEEauthorrefmark{1}Corresponding author: Jizhao Liu (email: liujz@lzu.edu.cn).
	}% <-this % stops a space
}

        % <-this % stops a space
%\thanks{This paper was produced by the IEEE Publication Technology Group. They are in Piscataway, NJ.}% <-this % stops a space
%\thanks{Manuscript received April 19, 2021; revised August 16, 2021.}}



\maketitle

\begin{abstract}
The rapid development of low-Earth orbit (LEO) satellite constellations and satellite communication systems has elevated the importance of secure video transmission, which is the key to applications such as remote sensing, disaster relief, and secure information exchange. In this context, three serious issues arise concerning real-time encryption of videos on satellite embedded devices: (a) the challenge of achieving real-time performance; (b) the limitations posed by the constrained computing performance of satellite payloads; and (c) the potential for excessive power consumption leading to overheating, thereby escalating safety risks. To overcome these challenges, this study introduced a novel approach for encrypting videos by employing two 1D chaotic maps, which was deployed on a satellite for the first time. The experiment on the satellite confirms that our scheme is suitable for complex satellite environments. In addition, the proposed chaotic maps were implemented on a Field Programmable Gate Array (FPGA) platform, and simulation results showed consistency with those obtained on a Raspberry Pi. Experiments on the Raspberry Pi 4B demonstrate exceptional real-time performance and low power consumption, validating both the hardware feasibility and the stability of our design. Rigorous statistical testing also confirms the scheme's resilience against a variety of attacks, underscoring its potential for secure, real-time data transmission in satellite communication systems.
\end{abstract}


\section{Introduction}

With the rapid advancement of satellite communication and the deployment of low-Earth orbit (LEO) satellite constellations, video data has become a critical information carrier in satellite communication systems, supporting applications such as remote sensing, disaster relief, and secure information exchange \cite{10254243}. However, the transmission of video data in satellite communication faces increasingly serious security threats, including information leakage, theft, and data distortion \cite{YANG20081299}. These security risks may not only endanger the functionality of the device and user privacy but also pose potential risks to national security. Consequently, the development of efficient and reliable encryption technology has become a core requirement for ensuring data security and privacy in satellite communications\cite{9787078}.

Existing video encryption algorithms can be broadly classified into two categories: traditional methods (e.g., DES, AES, and RSA) and chaos-based techniques \cite{jiang2024real}. Due to the complex mechanisms and the inherent redundancy in multimedia data, the effectiveness of traditional encryption in real-time satellite communication is hindered by limited encryption speed, particularly within the resource-constrained environments, such as portable satellite devices and LEO systems \cite{azzaz2020efficient}. Furthermore, conventional encryption methods often fail to preserve video storage formats, complicating real-time processing and playback \cite{hadjadj2022new}. The application of chaotic maps in cryptosystems has grown due to their sensitivity to initial conditions, randomness, and unpredictability, making them suitable for secure and adaptive video encryption in dynamic satellite communication environments \cite{10269762, 9782554, 9446980}. Due to these characteristics and the fast growth in both the practical implementation and theoretical understanding of chaos, a growing number of researchers have turned their attention towards the development of chaotic cryptography methods \cite{9340336, 9773990}. These developments underscore the potential of chaos-based approaches as a compelling alternative for addressing the unique security and performance challenges in satellite video transmission.

\begin{figure}[htbp]
	%	\hspace{-1cm} % 负值表示向左移动
	\centerline{\includegraphics[width=22pc]{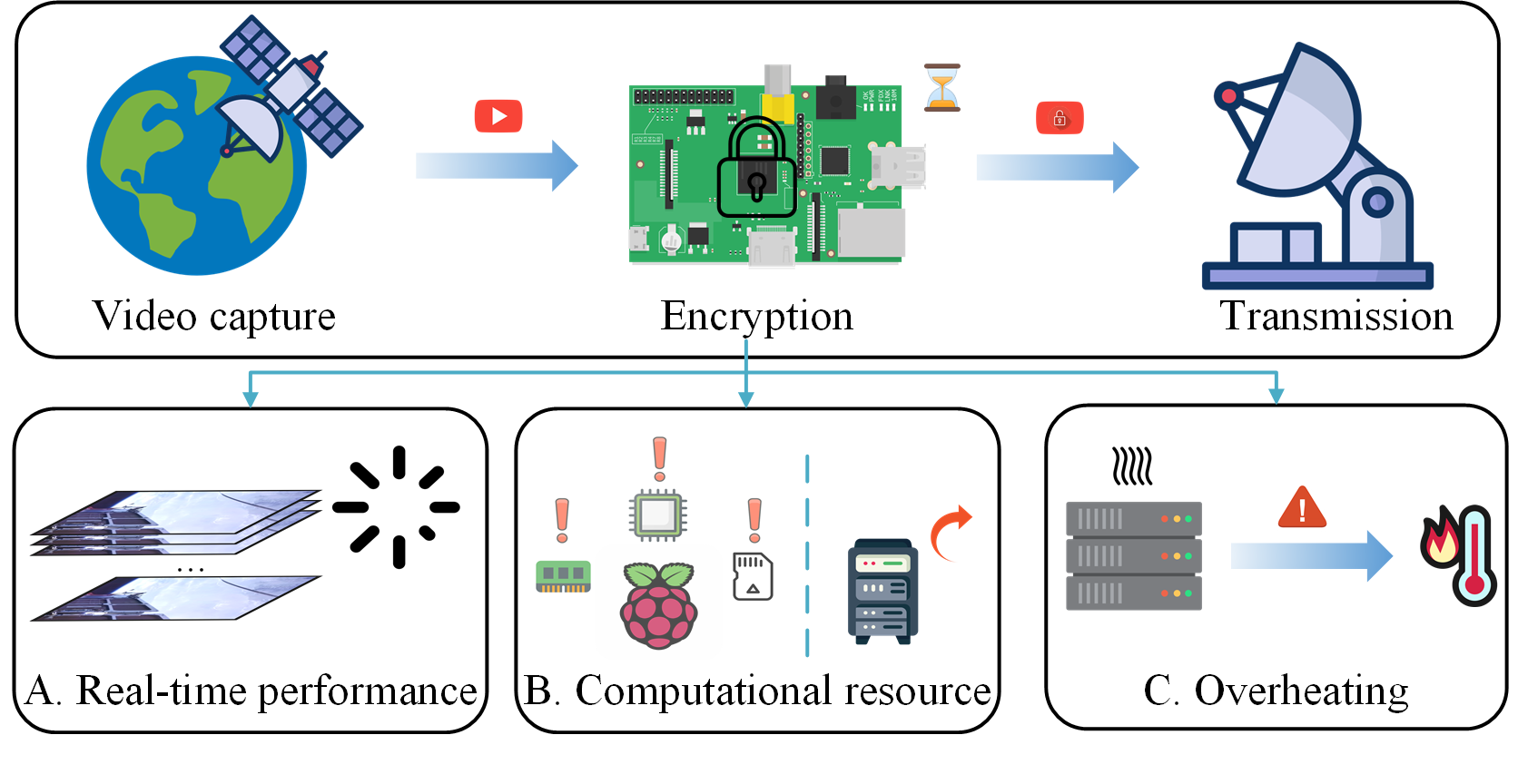}}
	\caption{Satellite video communication framework and challenges of real-time video encryption on satellite payloads.}
	\label{problems}
\end{figure}

In the context of satellite communication applications, the demand for secure video transmission on satellites is escalating, leading to a more widespread trend towards the adoption of video encryption. However, the deployment of encryption schemes on satellite payloads entails corresponding challenges: (a) Due to the complexity of encryption schemes, most video stream encryption approaches face challenges in achieving real-time characteristics. (b) The deployment of complex, high-performance equipment is hampered by the instability stemming from harsh environments. In comparison to ground-based high-performance computing devices, satellite devices generally possess fewer computing resources, thereby exacerbating challenges to the efficiency and stability of video encryption. (c) Due to the challenges of heat dissipation in a vacuum environment, the power required to operate the encryption algorithm must be confined within a specific range. Excessive power consumption may lead to overheating, resulting in equipment damage. However, overly high encryption speeds can also result in increased power consumption. Therefore, it is crucial to strike a balance between power consumption and encryption speed to ensure both system stability and performance. These challenges are shown in Fig. \ref{problems}.

%In order to achieve real-time video encryption on edge devices, our proposed encryption scheme utilizes a 1D chaotic system. Due to the simplicity and speed of the one-dimensional chaos map, the integration of this map into our encryption scheme enhances real-time performance. Furthermore, the minimal computational resources necessitated by 1D chaotic mapping make its implementation advantageous in low-performance edge devices. Additionally, given their exceptional unpredictability and sensitivity to initial values, chaotic maps offer robust security for cryptographic systems. Moreover, our method simplifies the encryption process to a solitary XOR operation, significantly diminishing computational complexity and encryption time. Finally, this solution saves the final encrypted data in a binary format. Storage of data in binary form minimizes the time required for storage. Additionally, the storage of encrypted data in text files ensures complete preservation of the data without any loss.

In this study, we proposed a novel video encryption algorithm based on 1D chaotic mapping and deployed it in a satellite environment for the first time. To overcome the three difficulties shown in Fig. \ref{problems}, four key strategies are employed: (a) Utilization of novel 1D chaotic systems. We integrate the 1D chaotic map, which is known for its simplicity and rapidity, into our encryption scheme to significantly boost real-time performance. (b) Implementation of the algorithm using C++. C++ is selected for its high performance and ability to handle computationally intensive operations with minimal latency. These characteristics make it ideal for real-time applications in resource-constrained satellite communication systems. (c) Simplification of the encryption process to a single XOR operation, which diminishes both computational complexity and encryption duration. The bitwise\_xor function in the OpenCV library further optimizes the encryption process by providing a highly efficient and hardware-accelerated operation for handling binary data, ensuring both speed and compatibility with embedded platforms. (d) Preservation of the final encrypted data in binary format. Binary storage substantially reduces the time necessitated for data storage and guarantees its intact retention without any loss. The proposed chaotic mapping was employed on a Field Programmable Gate Array (FPGA) platform, and simulation results showed consistent performance with those obtained on the Raspberry Pi platform, confirming its feasibility and stability across different satellite edge devices. Comprehensive statistical tests (NIST and DIEHARD) and various attack simulations further validate the robustness and security of our encryption system. Additionally, our timing analysis and power consumption analysis, which are conducted on the Raspberry Pi 4B, confirm the feasibility of real-time encryption of high-definition videos on a satellite payload, making it a promising solution for secure, real-time data transmission in satellite communication systems.

The main contributions of this work are as follows:

\begin{enumerate}
	\item We designed two novel and efficient 1D chaotic maps that are more effective in meeting real-time requirements in resource-constrained satellite communication systems compared to other chaotic maps. The lightweight bit shift operations in chaotic maps instead of floating-point arithmetic enhance the overall efficiency of the system. Additionally, due to their low-dimensional characters, we fully utilize the computation results from each iteration, maximizing efficiency and reducing redundancy in the process. The visual analysis and Poincare maps from the experiments confirm the complex chaotic behavior. Furthermore, this chaotic map successfully passed the DIEHARD test suite, demonstrating its high level of unpredictability.
	\item We introduced a novel video encryption scheme employing 1D chaotic mapping, including four important strategies: the adoption of novel 1D chaotic mapping, implementation using C++, the reduction of the encryption process to a single bitwise\_xor operation, and its storage in a binary file format. In addition, we considered two different application scenarios for satellite communication systems: (a) The first scenario involves the complete encryption process, including read, encrypt, and write operations, which is designed for applications that require end-to-end security. This application scenario is suited for the encryption of up to 1K resolution videos. (b) The second scenario focuses only on the encryption algorithm, analyzing whether the algorithm is suitable for real-time applications that are integrated into existing workflows. This application scenario can support video encryption for resolutions up to 2K.
	\item To the best of our knowledge, we first deployed a real-time video encryption algorithm on the satellite. The files and data transmitted from the satellite confirm that our algorithm can adapt to the complex space environment. This scheme successfully passed various image encryption tests, showcasing excellent real-time performance, low energy consumption, and compatibility with satellite embedded devices.
\end{enumerate}

The rest of this paper is organized as follows: Section \ref{secc} introduces the related work. We then propose the encryption scheme in Section \ref{sec3}. In Section \ref{sec4}, we discuss the performance and security analysis. Finally, the conclusions are drawn in Section \ref{sec6}.

\section{Related work}\label{secc}
%to this
%Relevant prior work includes studies of existing chaos-based encryption algorithms and video encryption schemes.

\subsection{Existing Chaos-Based Encryption Algorithms}
%Despite the widespread use of perfect ciphers like DES, IDEA, and RSA since the 1970s\cite{li2002chaotic}, their effectiveness in real-time multimedia environments is hindered by limited encryption speed. This is primarily due to the complex nature of encryption mechanisms and the inherent redundancy in multimedia data\cite{azzaz2020efficient}. In addition, conventional encryption methods like AES and DES have difficulties preserving storage formats, which adds complexity to real-time encryption efforts\cite{hadjadj2022new}. These issues underscore the significance of developing encryption techniques specifically designed to protect video streams.
%
%The application of chaotic maps in the construction of cryptosystems has grown due to their distinctive characteristics, such as their sensitivity to initial conditions, randomness, and unpredictability\cite{SETHI2022103746, zhang2024revealing, 9770397, 9782554}. Due to these characteristics and the fast growth in both the practical implementation and theoretical understanding of chaos, a growing number of researchers have turned their attention towards the development of chaotic cryptography methods\cite{9340336, Zhang_2024, 9773990}.

Chaotic maps may be categorized into two types: one-dimensional (1D) and high-dimensional chaotic maps. High-dimensional maps offer stronger security but are complex and time-consuming, making them inappropriate for real-time video encryption on satellite communication systems. In contrast, 1D chaotic maps, because of their uncomplicated structure, may be readily implemented, which is a very desirable characteristic in cryptography. Therefore, given the constraints of satellite communication systems, such as limited computational resources and the need for real-time processing, it is more appropriate to utilize 1D chaotic mapping for real-time video encryption.

Yang et al. presented a 2D map employed in the design of an S-box for image encryption, indicating high performance and robustness against various attacks \cite{e23101312}. Singh et al. proposed a novel encryption scheme based on chaotic systems and the DNA algorithm \cite{9779432}. Zhao et al. proposed a novel satellite image encryption algorithm that combines RNA computations with seven-dimensional complex chaotic systems, increasing security through scrambling and diffusion phases \cite{zhao2023satellite}.

% to
\subsection{Video Encryption Schemes}

Previous research has introduced a variety of encryption methodologies. Li et al. developed a novel encryption algorithm adapted to the H.264 standard \cite{li2005new}. Duan et al. introduced a CNN-based video encryption technique utilizing Faster RCNN \cite{duan2018novel}. Fadi et al. encrypted the video information through the RTP protocol \cite{almasalha2014partial}.

While these solutions offer high security due to numerous iterations and extensive calculations, they lack optimal efficiency and struggle to facilitate real-time encryption. Furthermore, the majority of experiments are conducted on computing devices equipped with high-performance processors, neglecting the consideration of limited computing resources in satellite communication systems. Particularly for performance-constrained satellite embedded systems, these encryption schemes present significant challenges to achieving real-time encryption.

\section{Proposed Method}\label{sec3}
\subsection{Problem Statement and Challenges}\label{sec21}
Achieving real-time video encryption is crucial for satellite communications. However, existing encryption algorithms struggle to deliver optimal real-time performance due to their substantial computational complexity, as depicted in Fig. \ref{problems}A. Furthermore, the harsh satellite environment and unstable energy supply hamper the deployment of high-performance satellite equipment, complicating the assurance of real-time performance and reliability for satellite communication systems. Additionally, the unique space environment makes heat dissipation a significant challenge, necessitating strict power range control for programs operating in satellite conditions and achieve a balance between power consumption and encryption speed. Thus, exploring real-time video encryption solutions that are suitable for satellite payloads becomes imperative to support the secure and resilient operation of satellite communication systems.

\subsection{Novel 1D Chaotic Map}

%In order to tackle the difficulties mentioned in Section \ref{sec21}, this study employs a discrete-space chaotic map to avoid the expensive integer division procedure. Ref. \cite{BEZERRA2023113160} uses discrete-space chaos mapping to avoid floating-point techniques, resulting in improved memory access efficiency. A representation of the chaotic map in a discrete space is shown below:
%
%%Due to the limited computational resources on satellites\cite{BENSIKADDOUR202050} and the real-time requirements for video transmission, it is necessary to choose a low-order chaotic system for encryption in order to reduce computational cost\cite{abbas2021hybrid}. Ref. \cite{BEZERRA2023113160} introduces a discrete-space chaotic map to avoid the costly integer division operation:
%

%
%While the initial chaotic map has low computational complexity and is appropriate for real-time video encryption, we have made modifications to transform the 2D discrete-space chaotic map into a novel 1D chaotic map. This alteration aims to further decrease the time required for generating pseudo-random sequences.

In order to tackle the difficulties mentioned in Section \ref{sec21}, this study employs a discrete-space chaotic map to avoid the expensive integer division procedure. Ref. \cite{BEZERRA2023113160} uses discrete-space chaos mapping to avoid floating-point techniques, resulting in improved memory access efficiency. The formula is shown in Eq. (\ref{chaos0}). While the initial chaotic map has low computational complexity and is appropriate for real-time video encryption, we have made modifications to transform the 2D discrete-space chaotic map into two novel 1D chaotic maps. This alteration aims to further decrease the time required for generating pseudo-random sequences while ensuring the efficient utilization of the generated sequences. The two novel chaotic maps are shown in Eq. (\ref{chaos}) and Eq. (\ref{chaos2}), respectively.

\begin{equation}
	\begin{cases} 
		a = \left(w \ll (w \% 64)\right) \, | \, \left(w \gg (64 - w \% 64)\right) \\
		w = \left(\frac{a}{2^{32}} + 1\right) \times \left(a \% 2^{32} + 1\right) + 1
	\end{cases}
	\label{chaos0}
\end{equation}

%to

\begin{equation}
	y_{k} = \begin{cases} 
		\left(y_{k-1} \ll (y_{k-1} \% 64)\right) \, | \\
		\, \left(y_{k-1} \gg (64 - y_{k-1} \% 64)\right) & \text{if } k \% 2 = 1 \\
		\left(\frac{y_{k-1}}{2^{32}} + 1\right) \times \\
		\left(y_{k-1} \% 2^{32} + 1\right) + 1 & \text{if } k \% 2 = 0
	\end{cases}
	\label{chaos}
\end{equation}

\begin{equation}
	y_{k} = \begin{cases} 
		\left(y_{k-1} \ll (y_{k-1} \% 64)\right) \, | \\
		\, \left(y_{k-1} \gg (64 - y_{k-1} \% 64)\right) & \text{if } k \% 2 = 0 \\
		\left(\frac{y_{k-1}}{2^{32}} + 1\right) \times \\
		\left(y_{k-1} \% 2^{32} + 1\right) + 1 & \text{if } k \% 2 = 1
	\end{cases}
	\label{chaos2}
\end{equation}

\begin{figure}[H]
	
	%	\hspace{-3cm} % 负值表示向左移动
	\centerline{\includegraphics[width=22pc]{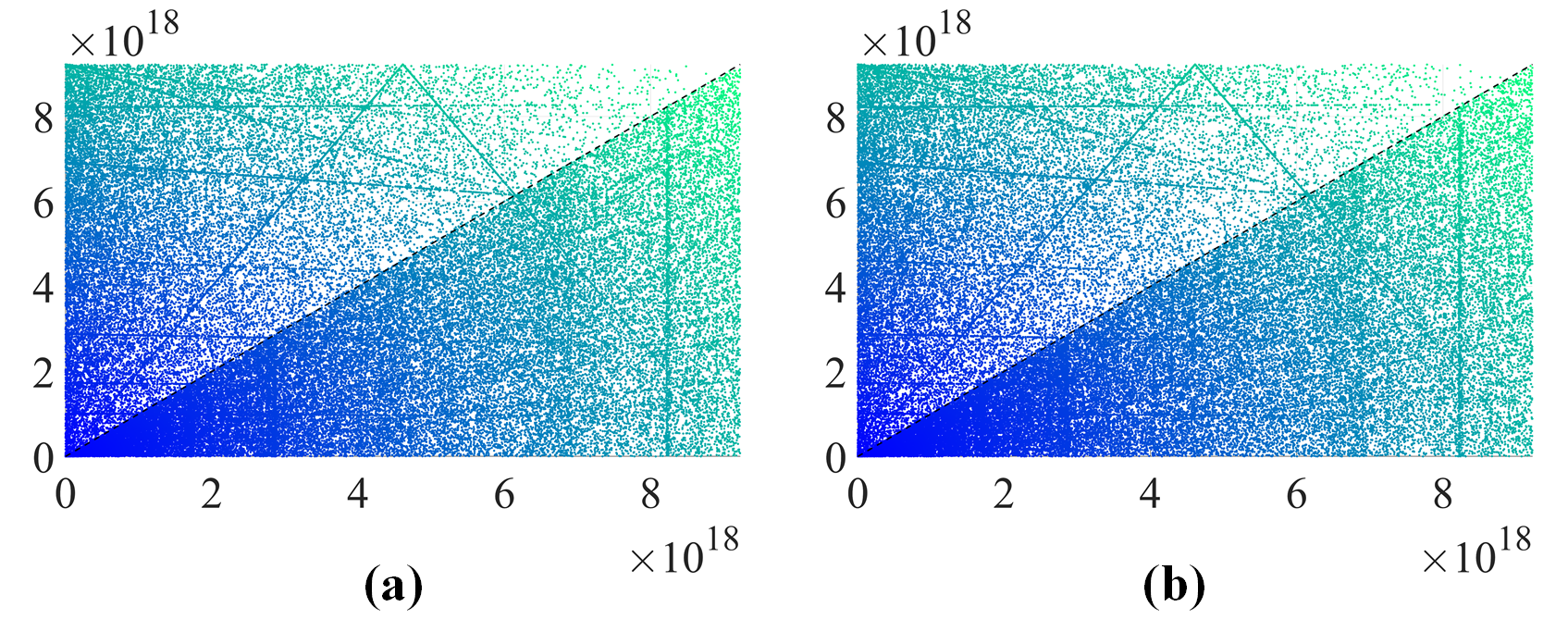}}
	\caption{Poincare section of the novel 1D chaotic map: (a) chaotic map (\ref{chaos}), (b) chaotic map (\ref{chaos2}).}
	\label{fig:111}
\end{figure}

The Poincare map is the point where a periodic orbit of a continuous dynamical system intersects a lower-dimensional subspace called the Poincare section, which is positioned in a way that it cuts across the flow of the system \cite{enwiki:1189115568}. Chaotic motion occurs when the Poincare section shows continuous curves or a large number of densely packed points \cite{XU2022313}. Fig. \ref{fig:111} provides evidence of the system's chaotic behavior.

\begin{figure}[H]
	
	%	\hspace{0.2cm} % 负值表示向左移动
	\centerline{\includegraphics[width=22pc]{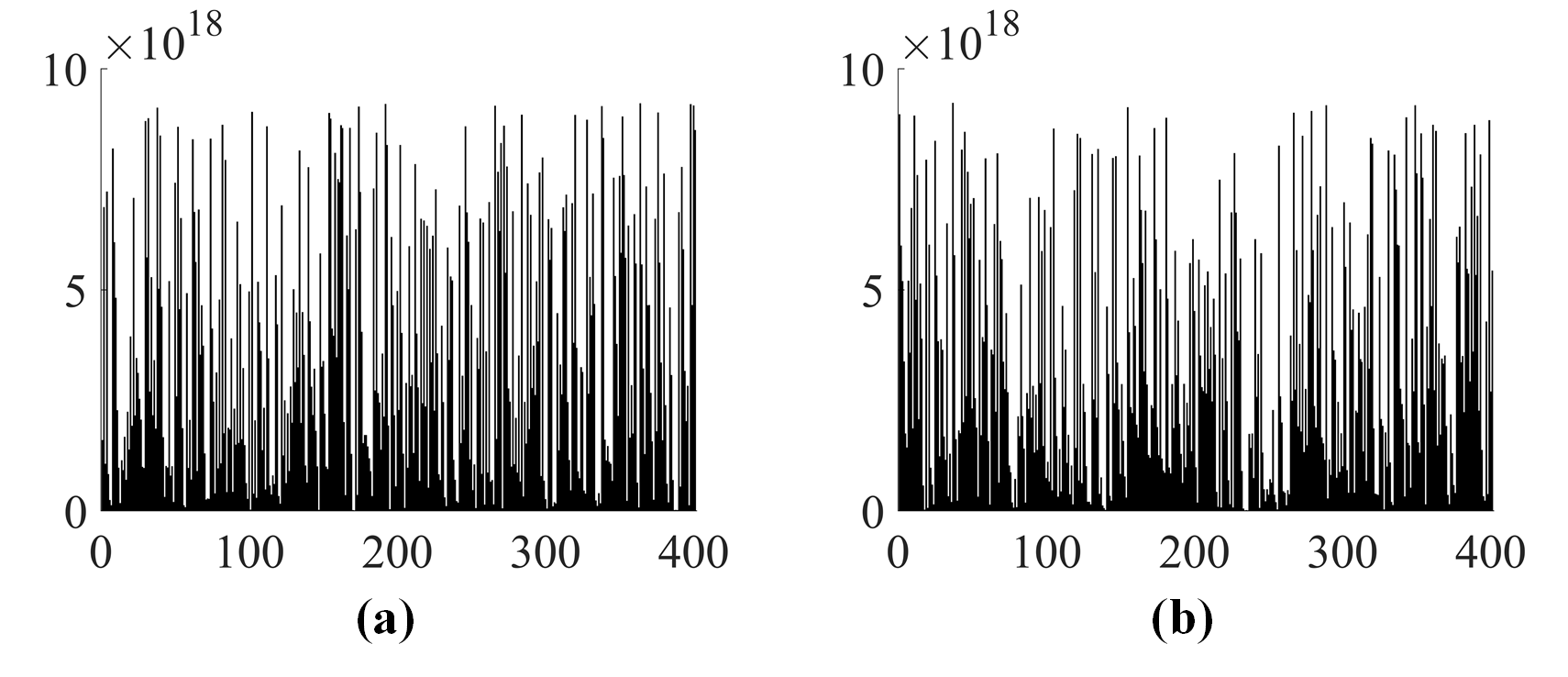}}
	\caption{Visual representation of novel 1D chaotic map: (a) chaotic map (\ref{chaos}), (b) chaotic map (\ref{chaos2}).}
	\label{fig:000}
\end{figure}

To assess the level of unpredictability in the pseudo-random sequence, the same initial values were applied to two chaotic maps, and the first 400 items were retrieved from each map. Fig. \ref{fig:000} illustrates each value within the first 400 elements of the two chaotic maps. It is shown that the difference between consecutive elements indicates exceptional randomness. Furthermore, the two sequences are entirely distinct, reflecting the inherent differences between the chaotic maps.

%\begin{figure}[H]
%	
%	%	\hspace{0.2cm} % 负值表示向左移动
%	\centerline{\includegraphics[width=16pc]{./generate_chaos_squence1.png}}
%	\caption{Visual representation of the novel 1D chaotic map \ref{chaos2}.}
%	\label{fig:0001}
%\end{figure}

The DIEHARD test suite consists of 17 distinct tests used to validate the randomness of a given sequence \cite{de2022secure}. A sequence is considered to have passed the test at a significance level of 0.05 if the resulting $P$ values lie within the range of 0.025 $<$ $P$ $<$ 0.975. Tab. \ref{die} demonstrates the pronounced randomness of two chaotic sequences. $(\checkmark)$ is used to indicate that the test was successfully passed.

%\begin{table}[H]
%	\centering
%	\setlength{\tabcolsep}{3pt}
%	\caption{Result of DIEHARD tests~suite.}
%	\label{die} 
%	\begin{tabular}{cccc}
	%		\toprule
	%		\textbf{Test} &  \textbf{\emph{p}-Value}& \textbf{\emph{p}-Value}\\
	%		\midrule
	%		Birthday spacing &0.8028 & PASSED \\
	%		
	%		Overlapping permutation & 0.5354 & PASSED \\
	%		
	%		Binary rank 32{$\times$}32 & 0.8732& PASSED  \\ 
	%		
	%		Binary rank 6{$\times$}8 & 0.9177& PASSED  \\
	%		
	%		Bitstream & 0.3148 & PASSED \\
	%		
	%		OPSO &  0.7322& PASSED \\
	%		
	%		OQSO &  0.8579& PASSED\\
	%		
	%		DNA &  0.7211 & PASSED\\
	%		
	%		Count the ones 01&0.1000 & PASSED\\
	%		
	%		Count the ones 02&0.8106& PASSED\\
	%		
	%		Parking lot&0.3937& PASSED\\
	%		
	%		2DS sphere&0.3486& PASSED \\
	%		
	%		3DS spheres&0.2722 & PASSED\\
	%		
	%		Squeeze&0.2749 & PASSED\\
	%		
	%		Overlapping sum&0.0566& PASSED \\
	%		
	%		Runs&0.7123& PASSED \\
	%		
	%		Craps&0.6480 & PASSED\\
	%		\bottomrule
	%	\end{tabular}
%\end{table}

\begin{table}[htbp]
	\centering
	\setlength{\tabcolsep}{3pt}
	\caption{Result of DIEHARD tests~suite.}
	\label{die} 
	\begin{tabular}{ccc}
		\toprule
		\textbf{Test} &  \textbf{Chaotic Map (\ref{chaos})}& \textbf{Chaotic Map (\ref{chaos2})}\\
		\midrule
		Birthday spacing &0.8028 (\checkmark)  & 0.6806 (\checkmark) \\
		
		Overlapping permutation & 0.5354 (\checkmark) & 0.7975 (\checkmark) \\
		
		Binary rank 32{$\times$}32 & 0.8732 (\checkmark)& 0.5353 (\checkmark)  \\ 
		
		Binary rank 6{$\times$}8 & 0.9177 (\checkmark)& 0.2626 (\checkmark)  \\
		
		Bitstream & 0.3148 (\checkmark) & 0.5024 (\checkmark) \\
		
		OPSO &  0.7322 (\checkmark)& 0.1155 (\checkmark) \\
		
		OQSO &  0.8579 (\checkmark)& 0.1203 (\checkmark)\\
		
		DNA &  0.7211 (\checkmark) & 0.5940 (\checkmark)\\
		
		Count the ones 01&0.1000 (\checkmark) & 0.9094 (\checkmark)\\
		
		Count the ones 02&0.8106 (\checkmark)& 0.2239 (\checkmark)\\
		
		Parking lot&0.3937 (\checkmark)& 0.8578 (\checkmark)\\
		
		2DS sphere&0.3486 (\checkmark)& 0.0445 (\checkmark) \\
		
		3DS spheres&0.2722 (\checkmark) & 0.5457 (\checkmark)\\
		
		Squeeze&0.2749 (\checkmark) & 0.9096 (\checkmark)\\
		
		Overlapping sum&0.0566 (\checkmark)& 0.0445 (\checkmark) \\
		
		Runs&0.7123 (\checkmark)& 0.3845 (\checkmark) \\
		
		Craps&0.6480 (\checkmark) & 0.5095 (\checkmark)\\
		\bottomrule
	\end{tabular}
\end{table}

\subsection{Pseudo-Random Sequence Generation}

Performing an XOR operation on an image with a pseudo-random sequence requires that each sequence value fall within the 0-255 range, corresponding to an 8-bit binary system. The pseudo-random sequence generation scheme introduced in this study reduces the iteration count of the chaotic sequence and consequently reduces the running time of sequence generation. The proposed scheme generates a single sequence by iterating only $\frac{m \times n}{8}$ times, resulting in a total of $\frac{m \times n}{4}$ iterations to generate the complete pseudo-random sequence.

\begin{enumerate}
	
	\item  Iterate two chaotic sequences of length $\frac{m \times n}{8}$, using the current timestamp as $y_0$, based on two novel 1D chaotic maps (\ref{chaos}) and (\ref{chaos2}), respectively.
	\item Each element of the chaotic sequence is treated as a 64-bit binary number and is subsequently divided into eight segments. These segments form the eight elements of the new pseudo-random sequence.
	%	\begin{equation}
		%		
		%		\begin{split}
			%			v_{8 \times j}     &= y_j[0: 7] \\
			%			v_{8 \times j + 1} &= y_j[8: 15] \\
			%			v_{8 \times j + 2} &= y_j[16: 23] \\
			%			v_{8 \times j + 3} &= y_j[24: 31] \\
			%			v_{8 \times j + 4} &= y_j[32: 39] \\
			%			v_{8 \times j + 5} &= y_j[40: 47] \\
			%			v_{8 \times j + 6} &= y_j[48: 55] \\
			%			v_{8 \times j + 7} &= y_j[56: 63]
			%		\end{split}
		%		\label{se}
		%	\end{equation}
\end{enumerate}

We utilize one of the chaotic systems to demonstrate the algorithm for generating pseudo-random sequences.

\begin{algorithm}
	\caption{Pseudo-Random Sequence Generation}
	\label{pse}
	\begin{algorithmic}[1]
		\Function{generate\_chaos\_sequence}{$m, n, t$}
		\State $ynext \gets t$
		\State $index \gets 0$
		\For{$i \gets 0$ to $\frac{m \times n}{8} - 1$}
		\For{$h \gets 0$ to 7}
		\State $Array[index] \gets ynext[8 \times h: 8 \times h + 7]$
		\State $index \gets index + 1$
		\EndFor
		\State $y \gets ynext$
		\If{$i \% 2 == 0$}
		\State $ynext \gets \left(\left(\frac{y}{2^{32}} + 1\right) \times \left((y \% 2^{32}) + 1\right) + 1\right)$
		\Else
		\State $ynext \gets (y \ll (y \% 64)) \mid (y \gg (64 - (y \% 64)))$
		\EndIf
		%		\State $ynext \gets y$
		%		\For{$h \gets 0$ to 7}
		%		\State $Array[index] \gets ynext[8 \times h: 8 \times h + 7]$
		%		\State $index \gets index + 1$
		%		\EndFor
		\EndFor
		\State \Return $Array$
		\EndFunction
	\end{algorithmic}
\end{algorithm}
\subsection{Encryption and Decryption Algorithm}

\begin{figure*}[htbp]
	
	%	\hspace{-1cm} % 负值表示向左移动
	\centerline{\includegraphics[width=34pc]{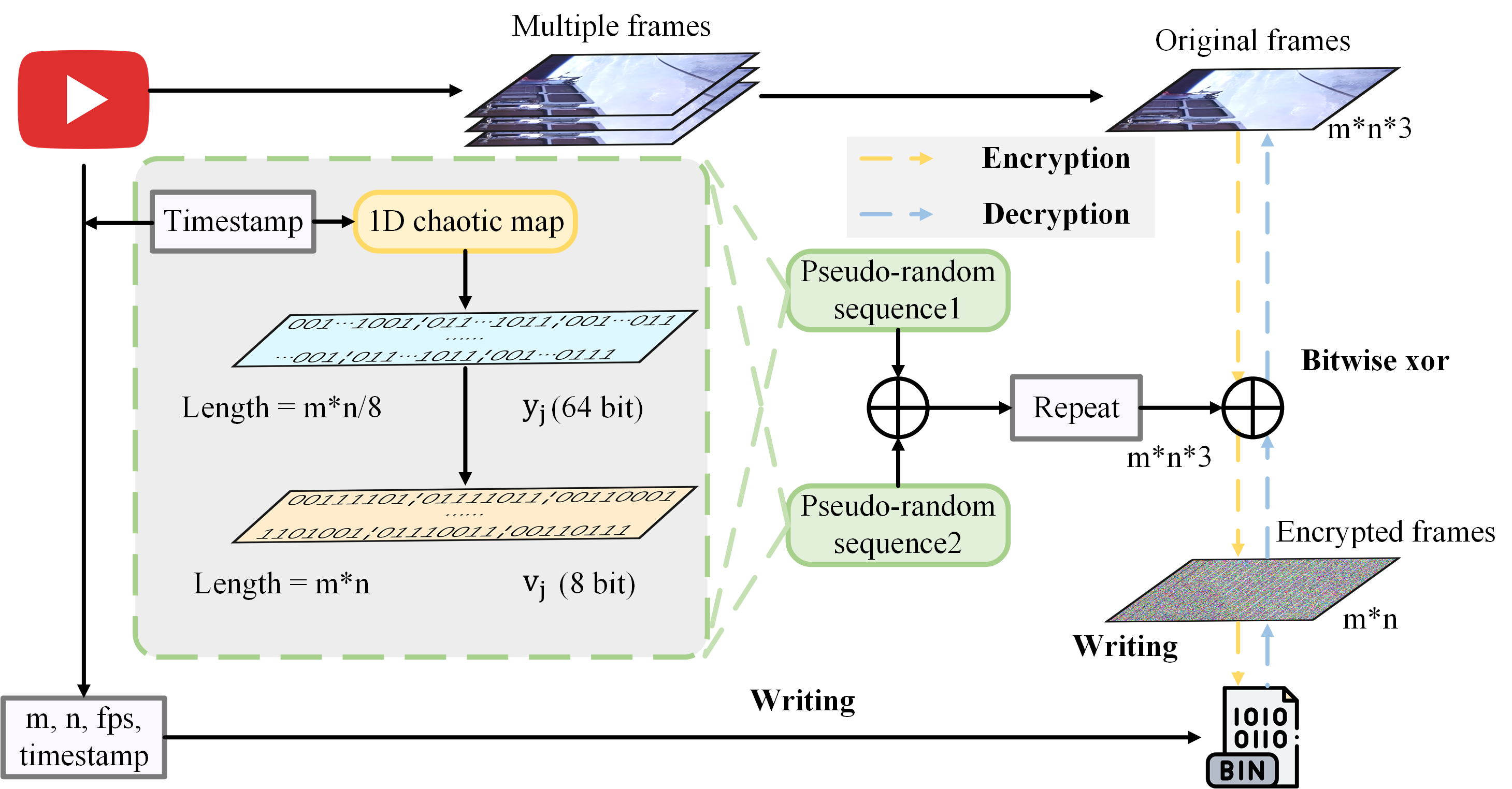}}
	\caption{Encryption process of the proposed algorithm.}
	\label{flowen}
\end{figure*}

To reduce computational complexity and decrease running time, we employ an encryption scheme based on novel 1D chaotic maps. As illustrated in Fig. \ref{flowen}, the scheme encompasses five stages: sequence generation, sequence processing, reading, XOR operation, and writing.

%The initial stage involves generating a pseudo-random sequence for XOR operations, based on the 1D chaotic map; the second stage is dedicated to extracting the image from each video frame; the third stage processes the image into an $m \times (3 \times n)$ size, enabling the simultaneous encryption of three channels; the fourth stage employs the image processed in the third stage for the XOR calculation; the fifth stage involves storing the encrypted image data in a bin file.

\begin{enumerate}
	
	\item Generate two pseudo-random sequences of length $m \times n$, utilizing novel 1D chaotic systems (\ref{chaos}) and (\ref{chaos2}).
	\item Execute an XOR operation between the two pseudo-random sequences.
	\item Expand the processed chaotic sequence into three channels and fill in an encrypted matrix pixel by pixel.
	\item Record the values of $m$, $n$, timestamp, and the frame rate of the video into a new binary file.
	\item Extract the original image from each frame of the video.
	\item Execute an XOR operation between the encrypted matrix and the original image.
	\item Store each pixel of the encrypted gray image into a binary file.
\end{enumerate}

\begin{algorithm}
	\caption{Encryption Algorithm}
	\label{Enc}
	Input: Original Video $(video)$
	
	Output: Encrypted File $(outFile)$
	\begin{algorithmic}[1]
		\State $m \gets \Call{get}{video, \texttt{FRAME\_HEIGHT}}$
		\State $n \gets \Call{get}{video, \texttt{FRAME\_WIDTH}}$
		\State $t \gets getCurrentTimestamp()$
		\State $c1 \gets \Call{generate\_chaos\_sequence1}{m, n, t}$
		\State $c2 \gets \Call{generate\_chaos\_sequence2}{m, n, t}$
		\State $c \gets \Call{bitwise\_xor}{c1, c2}$
		\State $En\_Mat \gets \Call{expandtorgb}{c}$
		\State $frameRate \gets \Call{get}{video, \texttt{FPS}}$
		\State \Call{write}{$outFile$, $m$, $n$, $frameRate$, $t$}
		\State $frameDataSize \gets m \times n$
		\While{\Call{read}{$video$, $frame$}}
		\State $En\_f \gets \Call{bitwise\_xor}{En\_Mat, frame}$
		\State \Call{write}{$outFile$, $En\_f.data$, $frameDataSize$}
		\EndWhile
		\State \Call{close}{$outFile$}
		\State \Call{release}{$video$}
	\end{algorithmic}
\end{algorithm}

The decryption scheme is essentially the inverse of the encryption scheme, and the decrypted frames are stored in the video.

%\begin{figure}[htbp]
%	
%	%	\hspace{-1cm} % 负值表示向左移动
%	\centerline{\includegraphics[width=22pc]{./flowde.png}}
%	\caption{Decryption process of the proposed algorithm.}
%	\label{flowde}
%\end{figure}

\begin{enumerate}
	
	\item Extract the video size, timestamp and video frame rate from the received binary file.
	\item Using the timestamp as the initial value, Generate two pseudo-random sequences of length $m \times n$, utilizing novel 1D chaotic systems (\ref{chaos}) and (\ref{chaos2}).
	\item Execute an XOR operation between the two pseudo-random sequences.
	\item Expand the processed chaotic sequence into three channels and fill in an decrypted matrix pixel by pixel.
	\item Retrieve each encrypted frame from the binary file.
	\item Execute an XOR operation between the decrypted matrix and the encrypted image.
	\item Encode each frame of the decrypted image into the decrypted video.
\end{enumerate}

%\begin{algorithm}
%	
%	\caption{Decryption Algorithm}
%	Input: Encrypted File $(inFile)$
%	
%	Output: Decrypted Video $(De\_video)$
%	\begin{algorithmic}[1]
	%		\State \Call{read}{$inFile$, $m$, $n$, $frameRate$, $Timestamp$}
	%		\State $w \gets \frac{n}{3}$
	%		\State $c, t \gets \Call{generate\_chaos\_sequence}{m, n, Timestamp}$
	%		\State $frameDataSize \gets m \times n$
	%		\State $De\_video \gets \Call{open}{\texttt{decrypted.mp4}, frameRate}$
	%		\While{\Call{read}{$inFile$, $restoredFrame$, $frameDataSize$}}
	%		\State $En\_Con\_f \gets \Call{bitwise\_xor}{restoredFrame, c}$
	%		\For{$i \gets 0$ \textbf{to} $2$}
	%		\State $channels[i] \gets En\_Con\_f[i \times w: i \times w + w, 0: m]$
	%		\EndFor
	%		\State $colorImage \gets \Call{merge}{channels[0: 2]}$
	%		
	%		\State \Call{write}{$De\_video$, $colorImage$}
	%		\EndWhile
	%		
	%		\State \Call{close}{$inFile$}
	%		\State \Call{release}{$De\_video$}
	%	\end{algorithmic}
%\end{algorithm}

\subsection{Satellite Video Communication Framework}

The upper part of Fig. \ref{problems} demonstrates the satellite video communication framework we adopted. Satellite video communication usually consists of three components: video capture, encryption, and data transmission. In this process, video data is captured by the satellite payload and generates a raw video stream. Subsequently, this video data is encrypted in real time by the proposed encryption scheme to ensure the security of the data during transmission. The encrypted data is saved as a binary file and then transmitted to the ground station through a satellite communication link.

\subsection{Experiments on a Satellite}

The experiment utilized a satellite payload on the Tiansuan Constellation platform, which is a Raspberry Pi 4B equipped with a 1.5GHz CPU. We tested the full time of running the encryption algorithm on the satellite and compared it with the experiment on the ground. The tests were conducted using two video samples: a 30s 360p video and a 10s 720p video, both recorded at 20 FPS. The experimental results are presented in Tab. \ref{p}.

\begin{figure}[htbp]
	\centering
	\includegraphics[width=18pc]{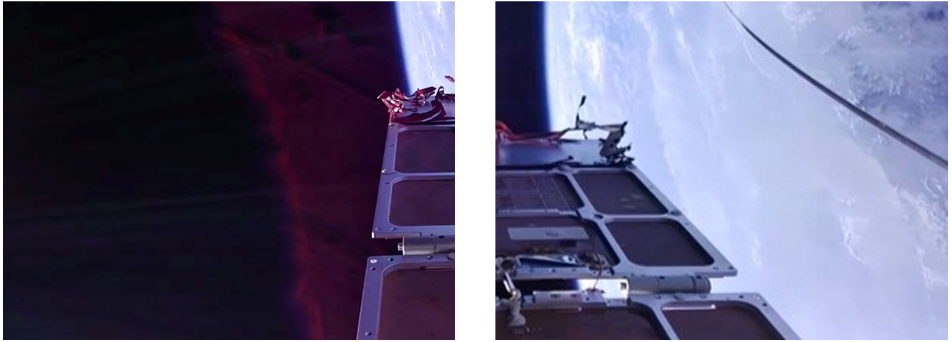}
	\caption{Cropped frames after decrypting binary file from satellite.}
	\label{frame}
\end{figure}

In addition, we obtained the encrypted binary file from the satellite for decryption and found that our encryption results did not lose data and could be completely restored to the original video. To protect the privacy of the brand in the advertisement, Fig. \ref{frame} displays cropped frames from the decrypted video. The experimental data on Tab. \ref{p} and Fig. \ref{frame} are provided by the Tiansuan Constellation platform.

It can be observed that the encryption time in each experiment is shorter than the duration of the original video, although the experimental data on the satellite shows slightly greater fluctuations compared to that on the ground. This may be attributed to temperature variations and power efficiency instability during satellite operation. Nevertheless, despite occasional performance instability, the experimental results remain largely consistent with those on the ground, demonstrating the capability for real-time encryption.

\begin{table}[htbp]
	\scriptsize
	\centering
	\setlength{\tabcolsep}{3pt}
	\caption{Comparison of experiments in space and on the ground.}
	\label{p}
	\begin{tabular}{ccccc}
		\toprule
		\textbf{Number of experiments}&\multicolumn{2}{c}{\textbf{Experiments in space}}&\multicolumn{2}{c}{\textbf{Experiments on the ground}}\\
		\cmidrule{2-5}
		& 360p & 720p & 360p & 720p\\
		\midrule
		1&2.2629 &2.8187&2.2534& 2.7189\\
		2&2.4708 &2.6597&2.2703&2.6462\\
		3&2.2378 &2.7568&2.3441&2.6731\\
		4&2.2432 &2.6597&2.2793&2.6983\\
		5&2.2670 &2.6727&2.2484&2.6766\\
		\midrule
		\textbf{95\% Interval}&2.2963&2.7135&2.2791&2.6827\\
		\textbf{Confidence}&$ \pm 0.0400$&$ \pm 0.0658$ &$ \pm 0.0338$&$\pm0.0239$\\
		\bottomrule
	\end{tabular}
	%	\vspace*{-0.5\baselineskip}
\end{table}

%\FloatBarrier

\subsection{FPGA Implementation}

Field Programmable Gate Array (FPGA) is a flexible programmable hardware platform with powerful parallel processing capabilities, which is suitable for implementing real-time and efficient digital signal processing and encryption algorithm applications. In order to ensure that our encryption scheme can be deployed on different edge devices, we deployed the proposed chaotic sequence generation scheme to Xilinx FPGA-Zynq xc020clg484. In this work, we used the Vivado 2018.3 version. First, we designed the IP core in the Vivado High-Level Synthesis (HLS) environment and imported the C++ code into the HLS project. To verify the correctness of the design, we also added a test bench file and used the Vivado HLS built-in simulator to perform a comprehensive functional simulation of the IP core. Subsequently, we perform synthesis and implement operations in sequence to convert the C++ code into a Verilog description and map the generated bitstream to the target FPGA device. Finally, the final deployment of the hardware circuit is achieved.

\begin{figure*}[htbp]
	\centering
	\includegraphics[width=32pc]{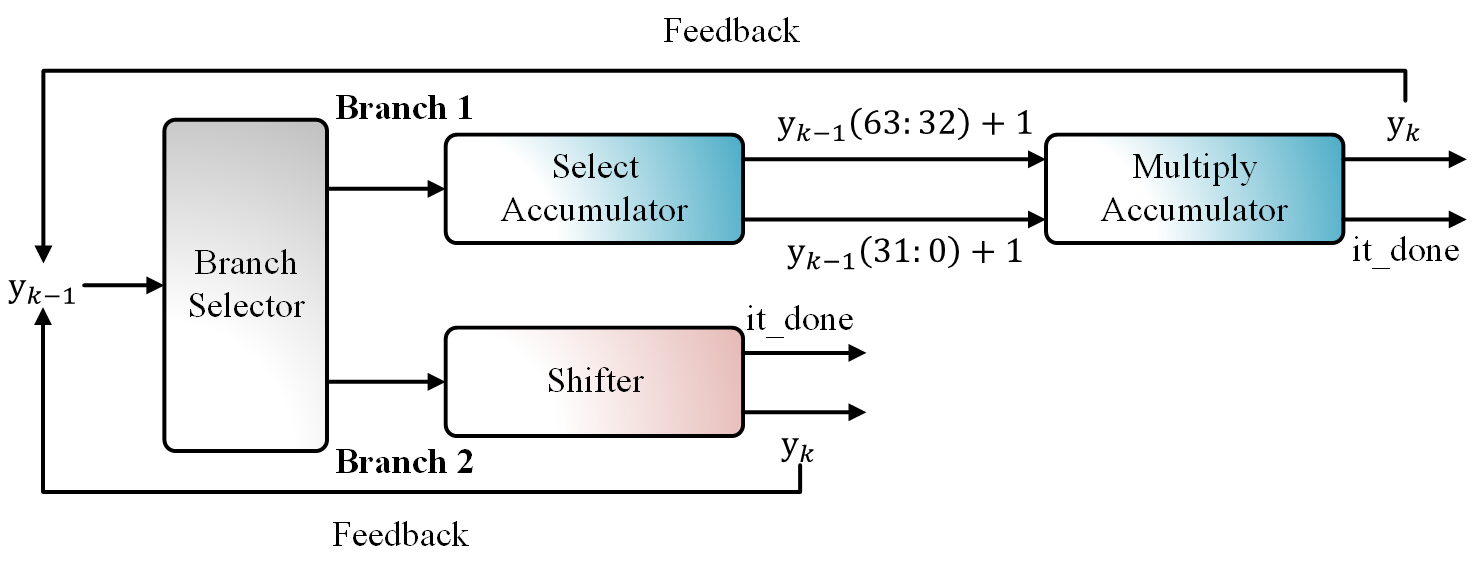}
	\caption{Block diagram of hardware architecture for the proposed chaotic map.}
	\label{design}
\end{figure*}

Fig. \ref{design} shows a simple block diagram of the chaotic map. In the hardware architecture of the chaotic map, the system has only one input $y_{k-1}$, which represents the state value of the previous iteration. The design first uses a Branch Selector module, which determines the data path according to the value of the current iteration index $k$. In Branch 1, the Select Accumulator module uses the registers and arithmetic units in the FPGA to process the upper 32 bits and lower 32 bits of the 64-bit input $y_{k-1}$, respectively, where $y_{k-1}(63:32) + 1$ corresponds to $\frac{y_{k-1}}{2^{32}} + 1$, and $y_{k-1}(31:0) + 1$ corresponds to $y_{k-1} \% 2^{32} + 1$. The design replaces division and remainder operations with shift operations, making full use of the efficient shift registers of the FPGA, thereby significantly improving the operation speed and resource utilization. In the next Multiply Accumulator module, the two outputs of the previous module are multiplied and added by 1 using a multiplier and an adder. On the other hand, in Branch 2, the Shifter module implements a circular shift on the 64-bit input $y_{k-1}$. The module first determines the shift amount by intercepting the lower 6 bits $y_{k-1}(5:0)$, then uses the left shift (shift amount) and right shift ($64 -$ shift amount) operations to split $y_{k-1}$ into two parts, and finally merges the two parts of the result through a bitwise OR operation. The system generates $y_{k}$ as the result of the current iteration of the chaotic map, which is saved by the register and feedback as the input $y_{k-1}$ of the next iteration. In addition, the status register generates an iteration completion signal $it\_done$ to indicate the end of this operation cycle.

\begin{figure*}[htbp]
	\centering
	\includegraphics[width=44pc]{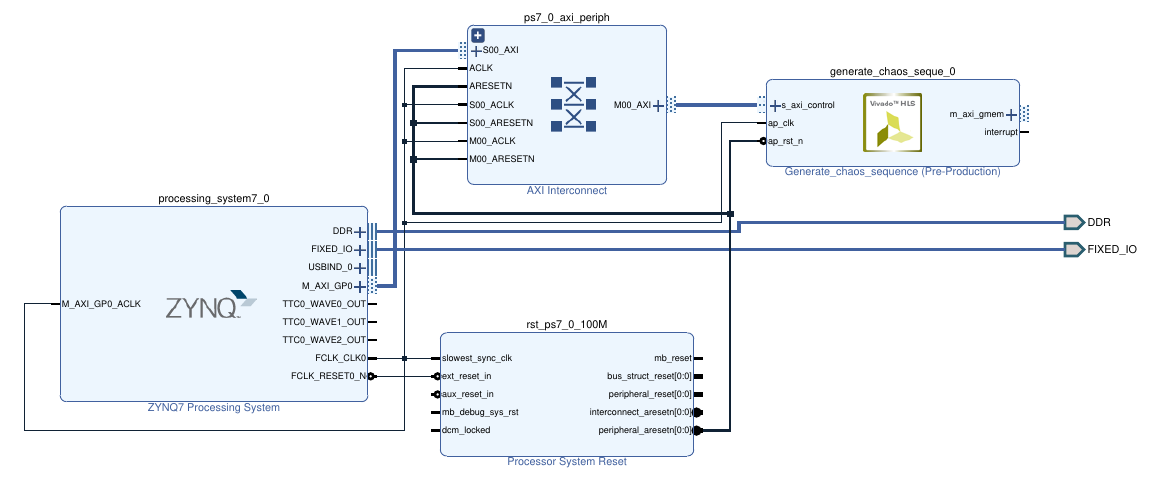}
	\caption{The architecture diagram through Vivado simulation.}
	\label{design_2}
\end{figure*}

Fig. \ref{design_2} shows the overall framework diagram.

%Fig. \ref{schematic} shows the RTL architecture of the scheme.

%\begin{figure}[htbp]
%	\centering
%	\includegraphics[width=32pc]{./schematic.pdf}
%	\caption{RTL presentation of the proposed scheme.}
%	\label{schematic}
%\end{figure}

Fig. \ref{com} shows the 10 simulation results of the proposed chaotic mapping in hexadecimal format on different edge devices. It can be seen that our chaotic system can achieve consistent simulation results on any hardware platform.

\begin{table}[htbp]
	\centering
	\setlength{\tabcolsep}{3pt}
	\caption{Comparison of simulation results for two chaotic maps on FPGA and Raspberry Pi platforms under identical initial values.}
	\label{com} 
	\begin{tabular}{cccc}
		\toprule
		\multicolumn{2}{c}{\textbf{Chaotic Map (\ref{chaos})}} & \multicolumn{2}{c}{\textbf{Chaotic Map (\ref{chaos2})}}\\
		Pi&FPGA&Pi&FPGA\\
		\midrule
		7598000000033e4a&7598000000033e4a&0844581288ce6a18&0844581288ce6a18\\
		00017d65438b3e4c&00017d65438b3e4c&1288ce6a18084458&1288ce6a18084458\\
		17d65438b3e4c000&17d65438b3e4c000&01bd6c9343bc2f34&01bd6c9343bc2f34\\
		10c029a7b4c5143a&10c029a7b4c5143a&f3401bd6c9343bc2&f3401bd6c9343bc2\\
		e84300a69ed31450&e84300a69ed31450&02baa010501ffd89&02baa010501ffd89\\
		0902c716f3b85529&0902c716f3b85529&754020a03ffb1205&754020a03ffb1205\\
		70aa5212058e2de7&70aa5212058e2de7&1d4dc628dea715c7&1d4dc628dea715c7\\
		0271e647f051b839&0271e647f051b839&a6e3146f538ae38e&a6e3146f538ae38e\\
		7204e3cc8fe0a370&7204e3cc8fe0a370&3c0832623be29e20&3c0832623be29e20\\
		4014c8524794147e&4014c8524794147e&3be29e203c083262&3be29e203c083262\\
		\bottomrule
	\end{tabular}
\end{table}

%It is evident that the encryption time for each experiment is shorter than the duration of the original video. Moreover, due to the limited computing resources and the unstable performance from repeated experiments, the efficiency of the satellite payload is not fully realized, leading to fluctuating performance. As indicated in Tab. \ref{p}, though certain experiments took significantly longer than others, all remained shorter in duration than the original video. The duration remains within an acceptable range.

%\section{Encryption and decryption algorithm}\label{sec3}
%The encryption system originates from a novel 1D chaotic map. An $m \times n$ color picture is separated into three channels: R, G, and B. The three channels are then converted into three gray images, subjected to further processing. Thereafter, these channels are concatenated to form a gray image of $m \times (3 \times n)$. Following this, the 1D chaotic equation is employed to generate a chaotic sequence, with each element being mapped to a 64-bit binary number. These elements are divided into 8 portions, each ranging from 0 to 256, to yield a new pseudo-random sequence. The gray image is then encrypted using XOR operation. Finally, the image is restored to an $m \times n$ color image by reversing the aforementioned steps, thereby completing the encryption process.
%
%
%\subsection{Encryption algorithm}
%
%The encryption algorithm is illustrated in Fig. \ref{flowen}.
%
%
%
%
%
%
%\subsection{Decryption algorithm}
%
%The decryption algorithm is depicted in Fig. \ref{flowde}.

\section{Performance and Security Analysis}\label{sec4}

To assess the effectiveness of real-time video encryption on satellite payloads, we conducted two types of analyses: (a) performance analysis and (b) security analysis. All experiments were conducted on a Raspberry Pi 4B, which is based on the Broadcom BCM2711 SoC and integrates a quad-core Cortex-A72 (ARM v8) 64-bit processor operating at 1.5GHz, 8GB LPDDR4-3200 SDRAM, and 64GB SD card storage, running a 64-bit Raspberry Pi OS.

\begin{figure*}[h!]
	
	\centerline{\includegraphics[width=26pc]{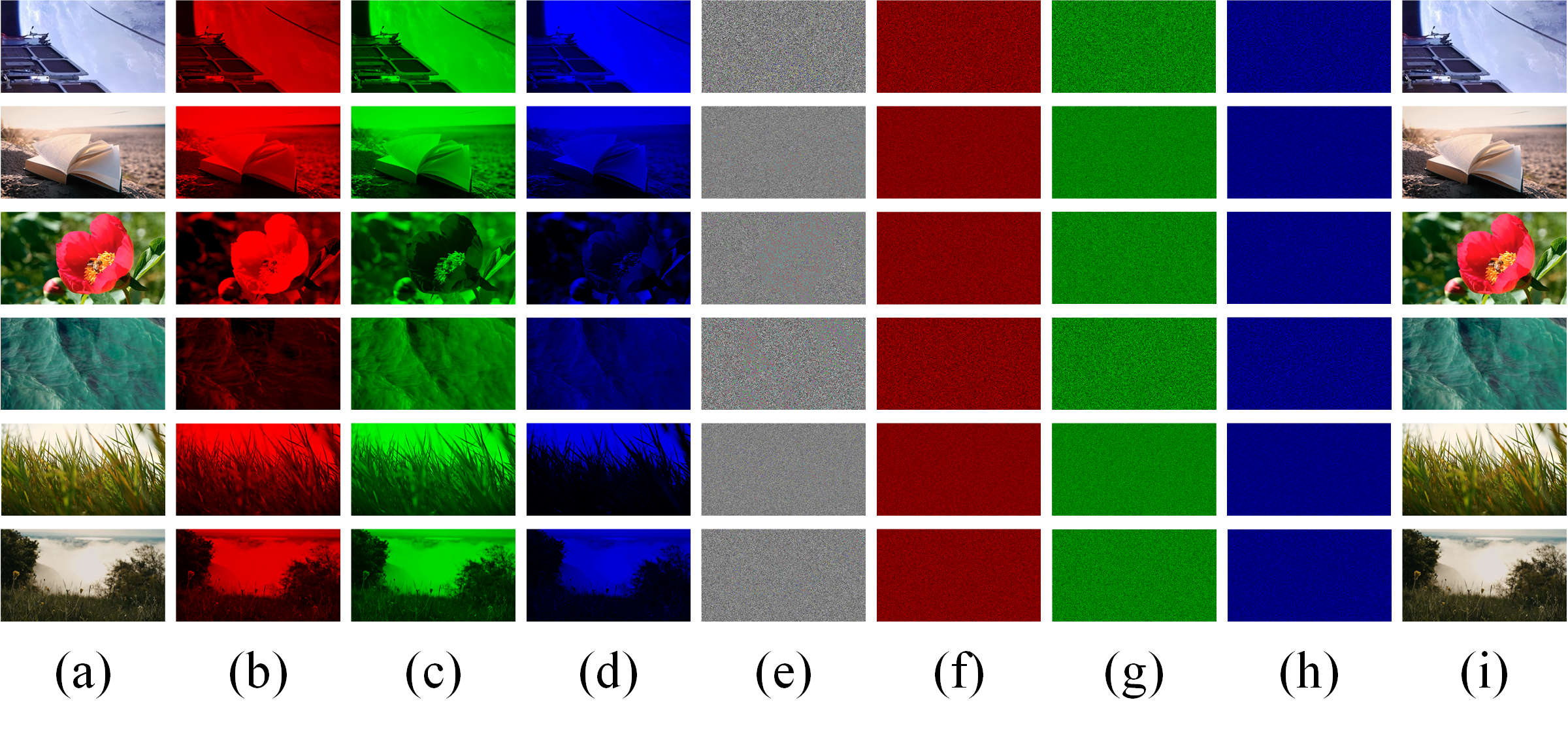}}
	\caption{Video frames: (\textbf{a},\textbf{b},\textbf{c},\textbf{d}) original images and their R, G, B channels, (\textbf{e},\textbf{f},\textbf{g},\textbf{h}) encrypted images and their R, G, B channels, (\textbf{i}) decrypted images.}
	\label{results}
\end{figure*}

%\subsection{Results on Raspberry Pi 4B}
%
Fig. \ref{results} displays randomly selected frames from the original, encrypted, and decrypted videos.

%\section{Performance and Security Analysis}\label{sec5}

\subsection{Time Analysis}

%The duration of the encryption process is a crucial parameter reflecting real-time performance of the scheme. Tab. \ref{entime} presents the encryption times for eight videos. 

We considered two application scenarios: (a) One scenario includes the complete encryption process, including five stages: sequence generation, sequence processing, reading, XOR operation, and writing. (b) The second scenario only considers the encryption algorithm. In our algorithm, the encryption process is only one XOR. Our design takes into account the requirement of different application scenarios, considering both the performance evaluation of the complete process to adapt to the performance bottleneck in practical applications and the high efficiency requirements in lightweight encryption scenarios.
%\begin{table}[h!]
%	\centering
%	\setlength{\tabcolsep}{3pt}
%	\caption{Encryption time of videos.}
%	\label{entime}
%	\begin{tabular}{cccccc}
	%		\toprule
	%		\textbf{Video}&\textbf{Width}&\textbf{Height}&\textbf{FPS}&\textbf{Original}&\textbf{Encryption}\\
	%		&&&&\textbf{video}&\textbf{time}\\
	%		&&&&\textbf{duration}&\\
	%		\midrule
	%		1 & 640 & 360 & 20 & 30 & 3.5607 \\
	%		2 & 640 & 360 & 29.97 & 13 & 2.5043 \\
	%		3 & 1280 & 720 & 29.97 & 13 & 8.9259 \\
	%		4 & 960 & 540 & 23.98 & 40 & 15.8698 \\
	%		5 & 960 & 540 & 30 & 30 & 14.5291 \\
	%		6 & 640 & 360 & 30 & 30 & 5.4055 \\
	%		7 & 1280 & 720 & 24 & 19 & 13.8626 \\
	%		8 & 960 & 540 & 25 & 18 & 5.8475 \\
	%		\bottomrule
	%	\end{tabular}
%\end{table}

%It is evident that the encryption times for high-definition videos are shorter than their original durations. Therefore, the encryption scheme demonstrates excellent real-time performance. 

%Tab. \ref{entime1} presents the encryption times for eight videos, comparing the proposed scheme with AES and demonstrating excellent real-time performance.
%
%Tab. \ref{entime1} presents a comparison between the proposed encryption scheme and the AES scheme. The results demonstrate that the processing time for the proposed encryption scheme is significantly shorter than that of AES algorithm, indicating that the real-time performance of our scheme is better than that of existing works.

Tab. \ref{entime1} presents the total encryption times and 95\% confidence intervals for eight videos. The results indicate that this scheme can encrypt up to 1K resolution videos while maintaining excellent real-time performance. During the experiments, it was observed that video reading and writing accounted for the majority of the processing time. To address this problem, we adopted a strategy of writing to binary files, significantly reducing write time and enhancing the real-time capability of the complete encryption process.

\begin{table}[htbp]
	\centering
	\setlength{\tabcolsep}{3pt}
	\caption{Encryption time of the complete encryption process.}
	\label{entime1}
	\begin{tabular}{cccccc}
		\toprule
		\textbf{Video}&\textbf{Width}&\textbf{Height}&\textbf{FPS}&\textbf{Original}&\textbf{Encrypted}\\
		&&&&\textbf{video duration}&\textbf{time}\\
		\midrule
		1 & 640 & 360 & 20 & 30& $2.3428 \pm 0.0924$\\
		2 & 640 & 360 & 29.97  & 13& $1.6247 \pm 0.0467$ \\
		3 & 1280 & 720 & 29.97  & 13& $5.4627 \pm 0.0568$\\
		4 & 960 & 540 & 23.98 & 40& $7.7184 \pm 0.0563$\\
		5 & 960 & 540 & 30 & 30& $6.9647 \pm 0.0470$\\
		6 & 640 & 360 & 30  & 30& $3.3859 \pm 0.0310$\\
		7 & 1280 & 720 & 24  & 19& $6.7034 \pm 0.0412$\\
		8 & 960 & 540 & 25  & 18& $3.6928 \pm 0.0657$\\
		9 & 1920 & 1080 & 29.97 & 11 & $10.9255 \pm 0.0715$\\
		\bottomrule
	\end{tabular}
\end{table}

Tab. \ref{entime2} lists the encryption times of the proposed algorithm and compares them with other algorithms. The bolded values in the table highlight the best results in terms of encryption time. The results show that, when considering only the encryption process, the proposed scheme can encrypt videos with resolutions up to 2K. Furthermore, compared to other algorithms, our method demonstrates faster encryption speeds and is more suitable for embedded devices with limited computational capabilities.

\begin{table*}[htbp]
	\centering
	\setlength{\tabcolsep}{3pt}
	\caption{Comparison of encryption algorithm time with different algorithms.}
	\resizebox{\textwidth}{!}{ % 将表格缩放到页面宽度
		\label{entime2}
		\begin{tabular}{ccccccccccc}
			\toprule
			\textbf{Video}&\textbf{Width}&\textbf{Height}&\textbf{FPS}&\textbf{Original}&\multicolumn{6}{c}{\textbf{Encrypted}}\\
			&&&&\textbf{video duration}&\multicolumn{6}{c}{\textbf{time}}\\
			\cmidrule{6-11}
			& & && &AES&Ref. \cite{rezaei2023secure} & Ref. \cite{BENTOUTOU2020176}&Ref. \cite{BEZERRA2023113160}& Ref. \cite{hafsa2022real}& Proposed\\
			\midrule
			7 & 1280 & 720 & 24  & 19&19.5309& 82.3405&20.9101&23.2585&40.3472 & \textbf{1.6955 $\pm$ 0.0147}\\
			8 & 960 & 540 & 25  & 18&10.6195& 69.9457&10.5545&12.6043&20.3263 &\textbf{0.8846 $\pm$ 0.0077}\\
			9 & 1920 & 1080 & 29.97 & 11 &30.4775& 138.847 &29.0570&36.3955&58.4864& \textbf{3.2532 $\pm$ 0.0741}\\
			10 & 1920 & 1020 & 24 & 26 &58.5893& 266.634 &57.0904 &70.4209&111.2180& \textbf{5.8099 $\pm$ 0.0198}\\
			11 & 2560 & 1440 & 30 & 30 &148.728& 713.092 &152.7620 &177.473&284.662& \textbf{15.3312 $\pm$ 0.3554}\\
			12 & 3840 & 2160 & 29.62 & 5 &57.4345& 321.101  &54.5954&68.7311&108.892& \textbf{5.2975 $\pm$ 0.0824}\\
			\bottomrule
		\end{tabular}
	}
\end{table*}

%Subsequently, we conduct a comprehensive analysis of the encryption time for a video. Tab. \ref{time} illustrates the total running time of the encryption scheme, and the running times for the five main phases of video 4.
%
%\begin{table}[htbp]
%	\centering
%	\setlength{\tabcolsep}{3pt}
%	\caption{Detailed analysis of encryption time.}
%	\label{time}
%	\begin{tabular}{cc}
	%		\toprule
	%		\textbf{Step}&\textbf{Time}\\
	%		\midrule
	%		Total time&   15.8698   \\                                
	%		Sequence generation&0.0239\\
	%		Reading&4.1067\\
	%		Image processing&1.7075\\
	%		XOR operation&1.3936\\
	%		Writing&8.8304\\
	%		\bottomrule
	%	\end{tabular}
%\end{table}
%
%The aggregate running time of these five main steps is very close to the total time, indicative of these steps being the primary time consumers. Further analysis revealed that the operations of reading video frames and writing files accounted for the majority of the time consumed, while generating a sequence occupied relatively little time.

\subsection{Power Consumption Analysis}

In the context of satellite communication systems, both the reliability and operational longevity of satellite devices are extremely critical. Given the high maintenance and energy supply costs associated with satellite payloads, conducting a power consumption analysis becomes necessary to ensure that the encryption algorithms employed do not surpass the allotted resource budget. Excessive power consumption not only threatens to reduce the operating life of the device but may also lead to overheating, thereby jeopardizing the security and real-time performance of the encryption algorithm in satellite communication applications.

\begin{figure}[htbp]
	%	\vspace{-0.2cm} % 负值表示向左移动
	%	\hspace{-0.15cm} % 负值表示向左移动
	\centerline{\includegraphics[width=22pc]{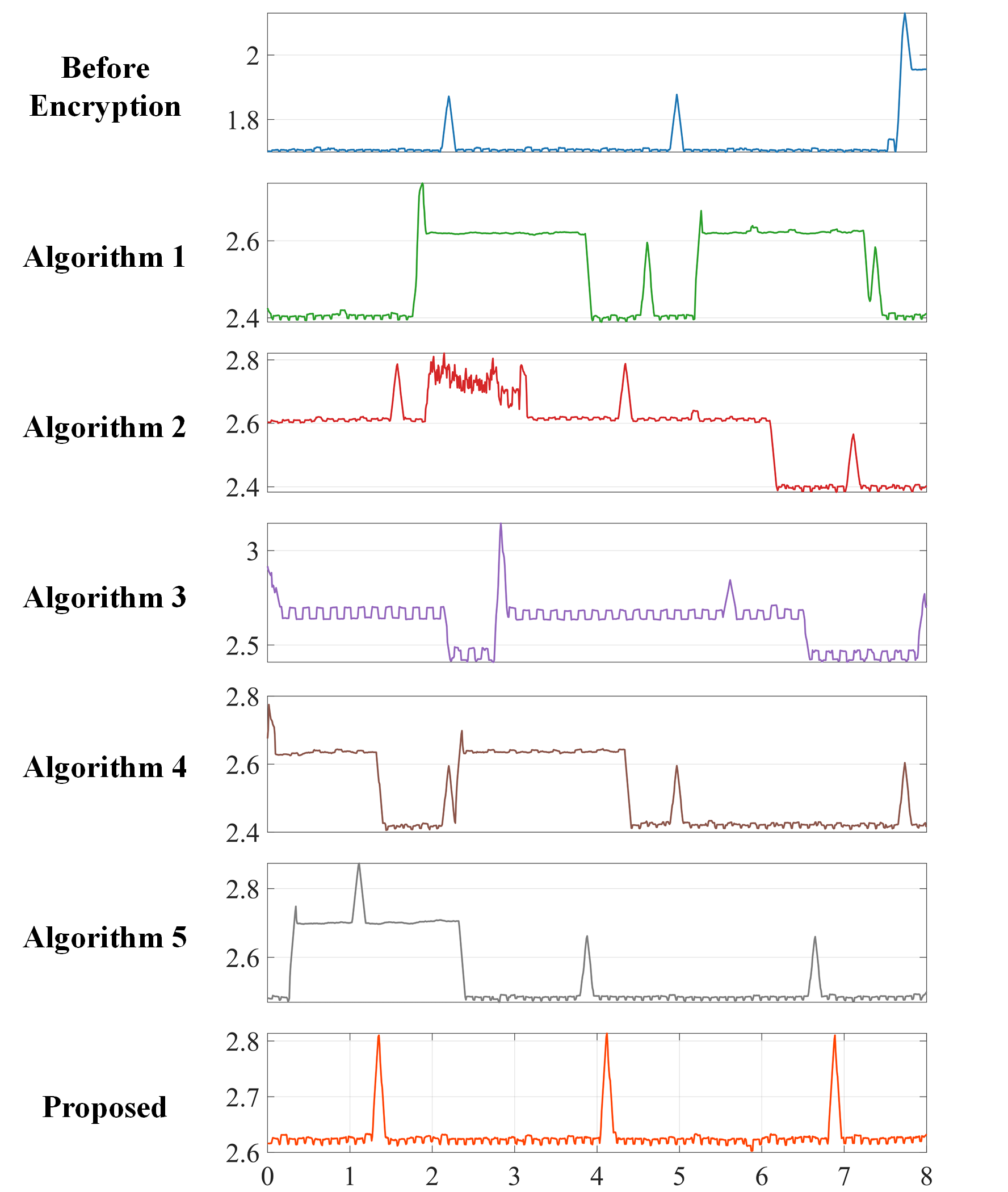}}
	\caption{Power (W) waveforms before and after the encryption scheme is executed, and the comparison with different algorithms.}
	\label{Power}
\end{figure}

To assess the power consumption characteristics of the proposed encryption algorithm, we employed the IoT Power CC device to measure power usage on a Raspberry Pi 4B, effectively simulating an embedded system environment representative of satellite payload constraints. Fig. \ref{Power} illustrates the power consumption before and during encryption, along with a comparison of the performance of other algorithms. Specifically, Algorithms 1, 2, 3, 4, and 5 correspond to Ref. \cite{rezaei2023secure}, Ref. \cite{BENTOUTOU2020176}, Ref. \cite{BEZERRA2023113160}, Ref. \cite{hafsa2022real}, and the AES algorithm, respectively. As shown, most algorithms maintain power consumption levels around 2.6W. However, due to slower encryption speeds, some algorithms occasionally exhibit lower power consumption, dropping to approximately 2.4W. In contrast, our proposed algorithm rarely demonstrates such lower power levels, attributed to its higher encryption efficiency. Overall, the power consumption of our algorithm remains stable and is not significantly higher than that of the other algorithms.

%The Raspberry Pi consumes nearly twice as much power during encryption compared to its idle state; however, the metric remains within acceptable limits.

%Tab. \ref{power} compares voltage, current, and power consumption for idle states and different videos encrypted using this algorithm. This table indicates that the video parameters have a low impact on power consumption, which falls within a reasonable range for each video.

%\begin{figure}[htbp]
%	
%	%	\hspace{-0.15cm} % 负值表示向左移动
%	\centerline{\includegraphics[width=18pc]{./Voltage.png}}
%	\caption{Voltage(V) waveforms before and after the encryption scheme is executed.}
%	\label{Voltage}
%\end{figure}
%
%\begin{figure}[htbp]
%	
%	%	\hspace{0.15cm} % 负值表示向左移动
%	\centerline{\includegraphics[width=18pc]{./Current.png}}
%	\caption{Current(mA) waveforms before and after the encryption scheme is executed.}
%	\label{Current}
%\end{figure}

%\begin{table}[htbp]
%	\centering
%	\setlength\tabcolsep{3pt}
%	\caption{Results of power consumption analysis.}
%	\label{power}
%	
%	\begin{tabular}{cccccccc}
	%		\toprule
	%		\textbf{Video}&\textbf{Width}&\textbf{Height}&\textbf{FPS}&\textbf{Original}&\textbf{Average }&\textbf{Average }&\textbf{Average }\\
	%		&&&&\textbf{video}&\textbf{voltage}&\textbf{current}&\textbf{power}\\
	%		&&&&\textbf{duration}&\textbf{(V)}&\textbf{(mA)}&\textbf{(W)}\\
	%		\midrule
	%		None & - & - & - & -  & 5.1078& 348.2856 & 1.7787\\
	%		1 & 640 & 360 & 20 & 30 & 5.0649&594.5605&3.0115 \\
	%		2 & 640 & 360 & 29.97 & 13  &5.0633&599.5931&3.0363\\
	%		3 & 1280 & 720 & 29.97 & 13 & 5.0564&635.8334&3.2148 \\
	%		4 & 960 & 540 & 540 & 40 & 5.0612&583.0637&2.9495\\
	%		5 & 960 & 540 & 30 & 30 & 5.0576&606.0333&3.0640 \\
	%		6 & 640 & 360 & 30 & 30 & 5.0604& 595.9326&3.0161 \\
	%		7 & 1280 & 720 & 24 & 19 & 5.0573&609.9392&3.0831 \\
	%		8 & 960 & 540 & 25 & 18 & 5.0582&607.8285&3.0744 \\
	%		\bottomrule
	%	\end{tabular}
%	
%\end{table}

\subsection{Key space analysis}

An effective encryption system requires a large key space that can successfully withstand attacks \cite{carlson2021evaluating}. According to cryptographic standards, the key space of a secure system should not be smaller than $2^{100}$ to effectively resist brute-force attacks \cite{QOBBI2023e01551}. Due to the 64-bit initial value $y_0$ of the proposed novel 1D chaotic map, the key space of the system is $2^{64} \times 2^{64} = 2^{128}$. This key space is large enough to withstand brute-force attacks.

\subsection{Histogram Analysis}

\begin{figure}[h]
	
	%	\hspace{-0.6cm} % 负值表示向左移动
	\centerline{\includegraphics[width=18pc]{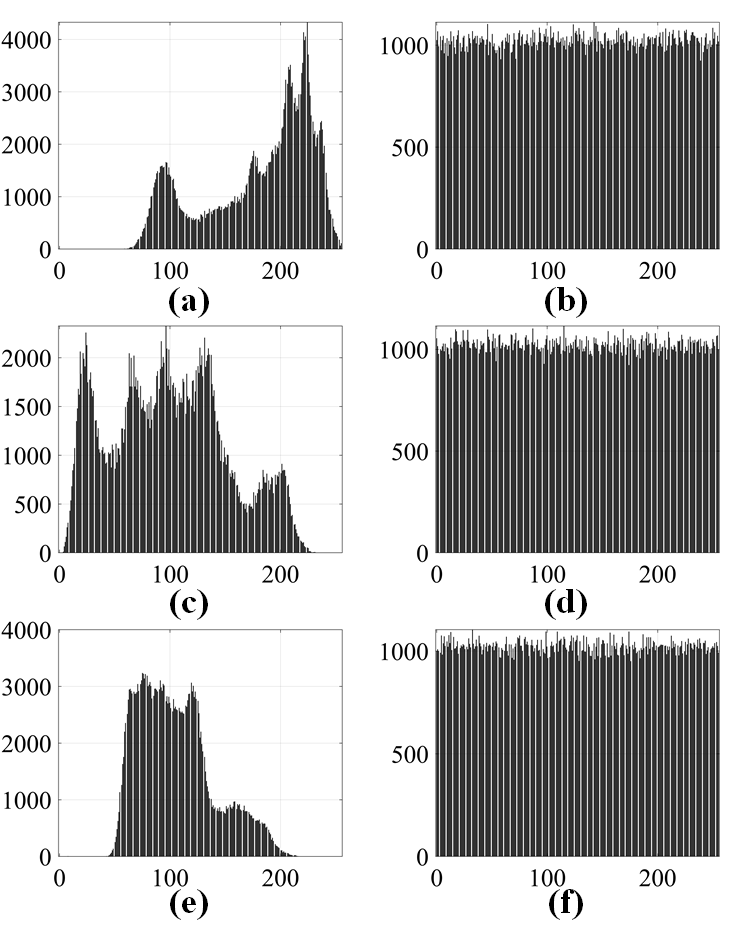}}
	\caption{Histograms of the original image and the encrypted image: (\textbf{a},\textbf{b}) R channel, (\textbf{c},\textbf{d}) G channel, (\textbf{e},\textbf{f}) B channel.}
	\label{his}
\end{figure}

Histogram analysis is employed to ascertain the distribution of pixel values in a particular image \cite{MALIK2020646}. Fig. \ref{his} displays histograms of the R, G, and B channels in both the original picture and the encrypted image. 

The dissimilarity between the histograms of the encrypted image and the original image is apparent. The encrypted histograms of the R, G, and B channels have a smooth distribution, suggesting excellent encryption performance.

\subsection{Information Entropy Analysis}

Entropy is a useful statistic for measuring the level of randomness. The information entropy approaches 8, indicating that the encryption technique is robust and secure \cite{10144593}. The calculation is performed as follows:

\begin{align}
	\text{H(k)} = -\sum_{i=0}^{2^n-1} p(k_j) \log_2 (p(k_j))
	\label{IE}
\end{align}

where the probability of the symbol $k_j$ is denoted by $p(k_j)$. Tab. \ref{H} shows information entropy for six images.

\begin{table}[htbp]
	\centering
	\setlength{\tabcolsep}{3pt}
	\caption{Results of information entropy analysis.}
	\label{H}
	\begin{tabular}{cccc}
		\toprule
		\textbf{Image}&	\textbf{Size}&	\textbf{Original images}&	\textbf{Encrypted images}\\
		\midrule
		1&	$640 \times 360$	&7.2723&	7.9997\\
		2&	$1280 \times 720$&	7.8131&	7.9999\\
		3&	$960 \times 540$&	7.4944&	7.9998\\
		4&	$640 \times 360$&	6.8443&	7.9997\\
		5&	$1280 \times 720$&	7.7089&	7.9999\\
		6&	$960 \times 540$&	7.4357&	7.9999\\
		\bottomrule
	\end{tabular}
\end{table}

The table shows that the information entropy closely aligns with the predicted value, suggesting that the technique is effective in defending against entropy attacks \cite{farah2020image}. Tab. \ref{H2} presents a comparison of the proposed encryption algorithm with other encryption algorithms for a $512 \times 512$ Lena image. The results show that the entropy achieved by our algorithm is closer to the ideal value, demonstrating superior performance.

\subsection{Correlation Analysis}

The correlation quantifies the association between adjacent pixels in the image \cite{ferdush2021chaotic}. Images with a high correlation are easier to crack, whereas images with a low correlation are more difficult to distinguish \cite{8955900}. The correlation coefficients are defined as:

\begin{equation}
	\begin{aligned}
		\left\{
		\begin{aligned}
			\text{$\mu(u)$} &= \frac{1}{M} \sum_{j=1}^{M} u_j \\
			\text{$\sigma^2(u)$} &= \frac{1}{M} \sum_{j=1}^{M} (u_j - \mu(u))^2 \\
			\text{$\kappa(u, v)$} &= \frac{1}{M} \sum_{j=1}^{M} (u_j - \mu(u))(v_j - \mu(v)) \\
			\text{$\rho(u, v)$} &= \frac{\kappa(u, v)}{\sigma(u) \sigma(v)}
		\end{aligned}
		\right.
	\end{aligned}
\end{equation}

\begin{table}[htbp]
	\centering
	\setlength{\tabcolsep}{3pt}
	\caption{Comparison of information entropy with different algorithms.}
	\label{H2}
	\begin{tabular}{cc}
		\toprule
		\textbf{Algorithm}&	\textbf{Entropy}\\
		\midrule
		Ref. \cite{rezaei2023secure} & 7.9994 \\
		Ref. \cite{BENTOUTOU2020176}& 7.9994 \\
		Ref. \cite{BEZERRA2023113160}& 7.9991 \\
		Ref. \cite{zhao2023satellite}&7.9985 \\
		Ref. \cite{mao2023real} & 7.9977 \\
		Ref. \cite{LI2022111989} & 7.9975\\
		Ref. \cite{SU2024121088} & 7.9993 \\
		Ref. \cite{UMAR2024125050} & 7.9994\\
		Proposed & \textbf{7.9998} \\
		\bottomrule
	\end{tabular}
\end{table}

where $u$ and $v$ represent two adjacent pixels in the horizontal, vertical, and diagonal directions. The correlation analysis results for six images are presented in Tab. \ref{corr}, which shows the pixel value correlation.

\begin{table}[htbp]
	\centering
	\setlength{\tabcolsep}{2pt}
	\caption{Results of correlation analysis.}
	\label{corr}
	\begin{tabular}{ccccccccc}
		\toprule
		\textbf{Image}&	\textbf{Size}&\textbf{Channels}&\multicolumn{3}{c}{\textbf{Original images}}&\multicolumn{3}{c}{\textbf{Encrypted images}}\\
		\cmidrule{4-9}
		& & & Horiz. & Vert. & Diag. & Horiz. & Vert. & Diag. \\
		\midrule
		& $640$ & R &  0.9924 &  0.9916 &  0.9890 & -0.0035 &  0.0033 & -0.0042 \\
		1&$\times$ & G & 0.9943 &  0.9936 &  0.9917 & -0.0018 &  0.0035 & -0.0039 \\
		&$360$ & B & 0.9949 &  0.9940 &  0.9923 & -0.0040 &  0.0019 &  0.0005 \\
		& $1280$ & R & 0.9987 & 0.9971 & 0.9963 & 0.0054 &  0.0023 &  0.0006 \\
		2&$\times$ & G & 0.9988 & 0.9971 & 0.9965 & 0.0045 &  0.0024 &  0.0007 \\
		&$720$ & B & 0.9986 & 0.9967 & 0.9961 & 0.0040 &  0.0034 &  0.0011 \\
		& $960$ & R & 0.9966 & 0.9970 & 0.9933 & -0.0013 &  0.0027 & -0.0005 \\
		3&$\times$ & G & 0.9910 & 0.9911 & 0.9820 & -0.0008 &  0.0041 &  0.0003 \\
		&$540$ & B & 0.9887 & 0.9887 & 0.9766 & -0.0004 &  0.0038 & -0.0008 \\
		& $640$ & R & 0.9799 & 0.9752 & 0.9657 & 0.0039 &  0.0042 &  0.0018 \\
		4&$\times$ & G & 0.9856 & 0.9841 & 0.9770 & 0.0024 &  0.0027 &  0.0006 \\
		&$360$ & B & 0.9791 & 0.9763 & 0.9656 & 0.0014 &  0.0027 &  0.0037 \\
		& $1280$ & R & 0.9828 & 0.9949 & 0.9881 & -0.0002 &  0.0029 & -0.0024 \\
		5&$\times$ & G & 0.9815 & 0.9945 & 0.9873 & -0.0010 &  0.0032 & -0.0016 \\
		&$720$ & B & 0.9861 & 0.9958 & 0.9902 & 0.0001 &  0.0031 & -0.0017 \\
		& $960$ & R & 0.9954 & 0.9977 & 0.9935 & -0.0010 &  0.0027 & -0.0008 \\
		6&$\times$ & G & 0.9952 & 0.9977 & 0.9934 & -0.0005 &  0.0023 & -0.0001 \\
		&$540$ & B & 0.9955 & 0.9978 & 0.9938 & -0.0002 &  0.0036 &  0.00002 \\
		\bottomrule
	\end{tabular}
\end{table}

The correlations in the proposed method are noticeably lower compared to the original images \cite{YILDIRIM2021166728}. Tab. \ref{corr2} compares the correlation results of the Lena image encrypted by different algorithms and the proposed algorithm. It can be observed that our encryption algorithm achieves relatively low correlation values across all three directions, with three indices being the lowest among the comparisons.

\begin{table}[htbp]
	\centering
	\setlength{\tabcolsep}{2pt}
	\caption{Comparison of correlation with different algorithms.}
	\label{corr2}
	\begin{tabular}{ccccc}
		\toprule
		\textbf{Algorithm}&\textbf{Channels}&\multicolumn{3}{c}{\textbf{Correlation}}\\
		\cmidrule{3-5}
		&& Horiz. & Vert. & Diag.\\
		\midrule
		&R&0.0031&	0.0080&0.0043\\
		Ref. \cite{rezaei2023secure} &G& -0.0029 & -0.00002 & \textbf{-0.0014} \\
		&B&0.0035&-0.0057 & \textbf{0.0004} \\
		&R&-0.0032&\textbf{-0.0009}&-0.0031\\
		Ref. \cite{BEZERRA2023113160}&G&0.0044&\textbf{-0.0003}&0.0019 \\
		&B&-0.0034 &-0.0008& 0.0027 \\
		&R&0.0011&-0.0098&-0.0071\\
		Ref. \cite{SHENG2023103592}&G& \textbf{-0.0025}&	-0.0120&-0.0090 \\
		&B&	0.0003&	-0.0082&-0.0065\\
		&R&-0.0050 & -0.0096&0.0018\\
		Ref. \cite{WANG2022107753}&G& \textbf{0.0025} & -0.0032 & 0.0015 \\
		&B&	0.0035 & -0.0023 & -0.0042\\
		&R&-0.0045 & 0.0149 & -0.0033\\
		Ref. \cite{JASRA2022117861}&G& 0.0026 & 0.0126 & -0.0013 \\
		&B&	\textbf{-0.0001} &0.0074 & 0.0021\\
		&R&\textbf{-0.0004} &  0.0014 &\textbf{ -0.0003}\\
		Proposed &G& -0.0050 & -0.0006 & -0.0032 \\
		&B&-0.0032 & \textbf{-0.0006} & -0.0006 \\
		\bottomrule
	\end{tabular}
\end{table}

\begin{figure}[htbp]
	%	\vspace{-0.4cm} % 负值表示向左移动
	%	\hspace{-0.2cm} % 负值表示向左移动
	\centerline{\includegraphics[width=22pc]{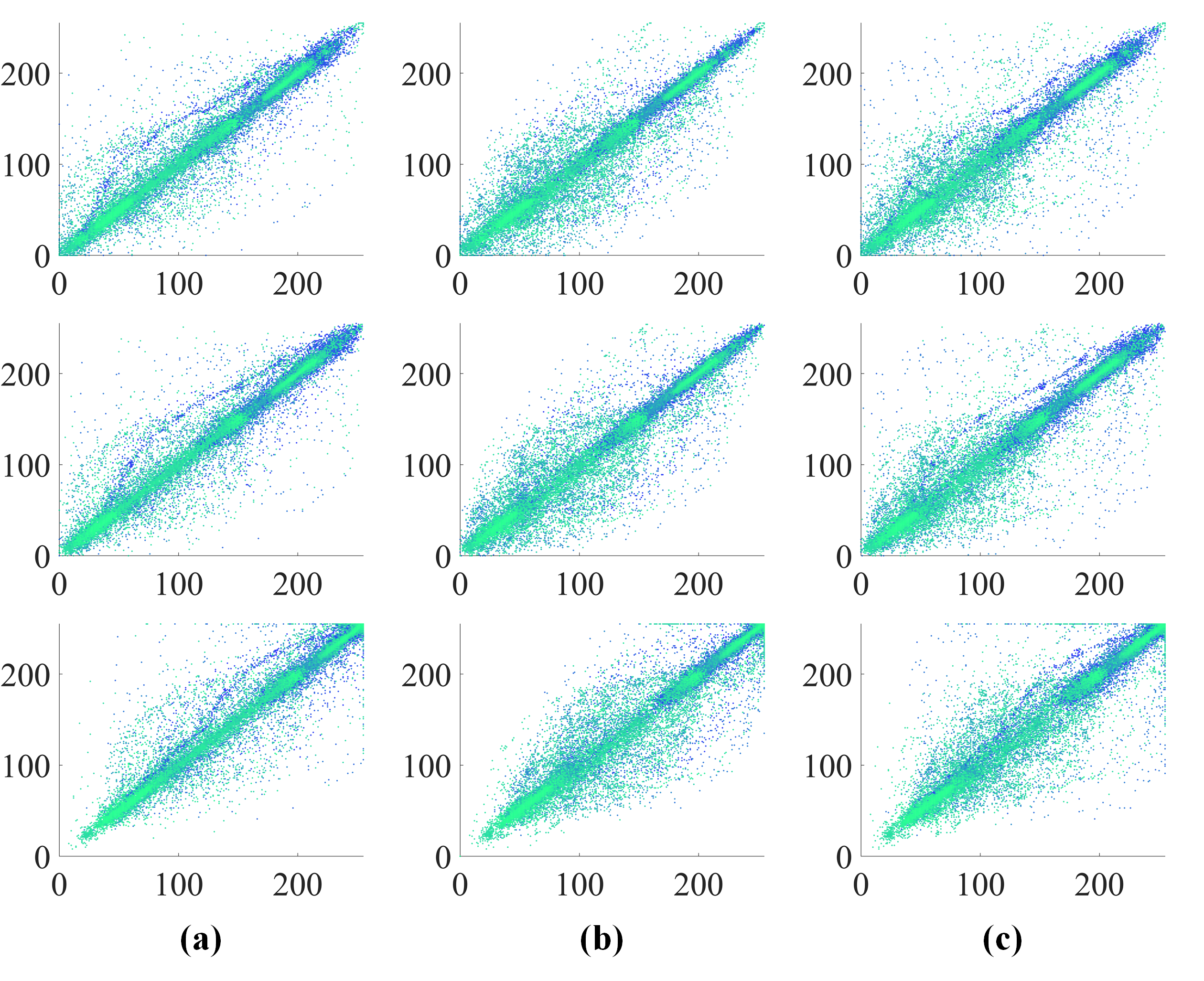}}
	\caption{Correlation analysis of the original images of the R, G, B channels: (\textbf{a}) horizontal direction, (\textbf{b}) vertical direction, (\textbf{c}) diagonal direction.}
	\label{orcorr}
\end{figure}

\begin{figure}[htbp]
	
	%	\vspace{-0.4cm} % 负值表示向左移动
	%	\hspace{0.2cm} % 负值表示向左移动
	\centerline{\includegraphics[width=22pc]{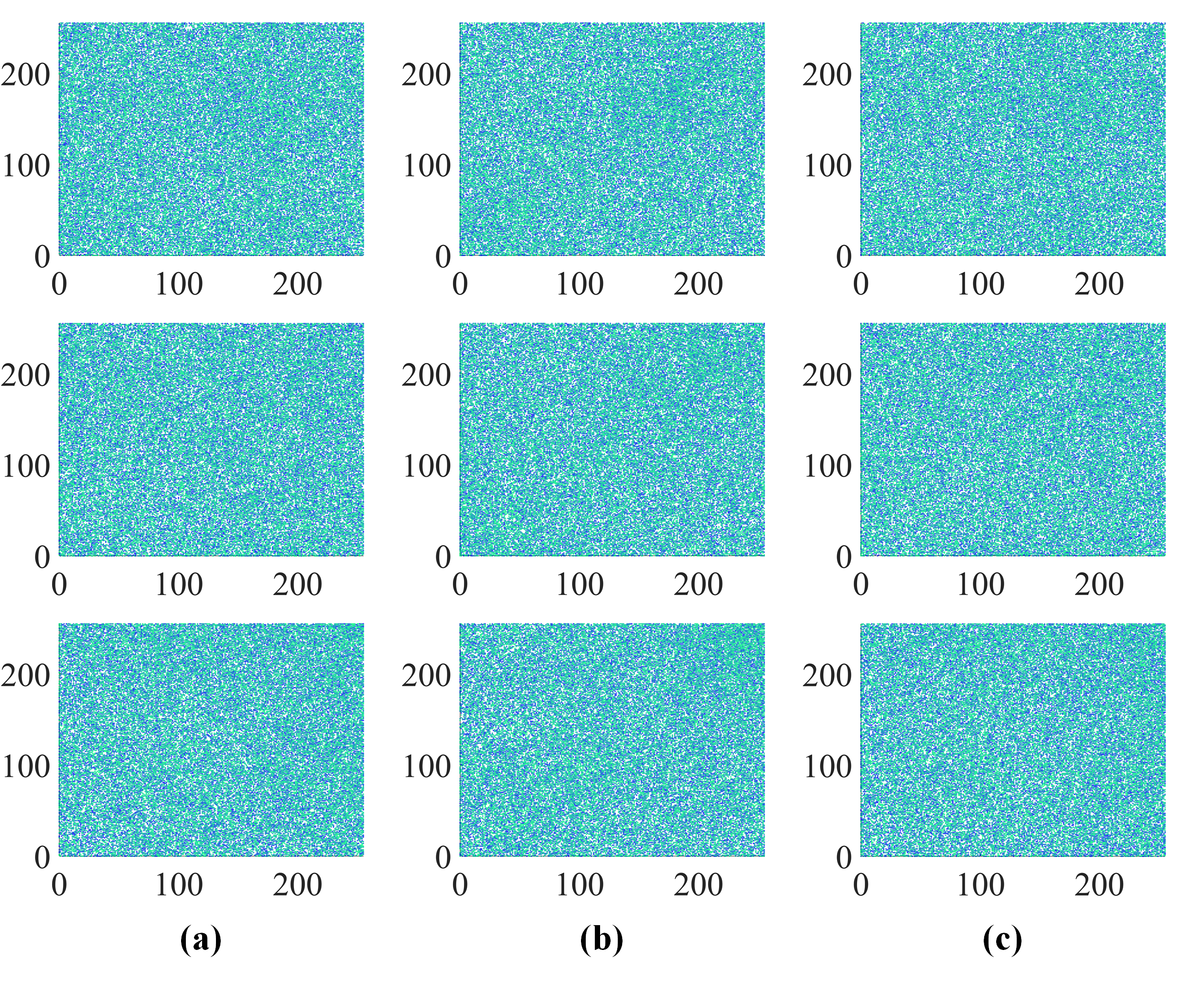}}
	\caption{Correlation analysis of the encrypted images of the R, G, B channels: (\textbf{a}) horizontal direction, (\textbf{b}) vertical direction, (\textbf{c}) diagonal direction.}
	\label{encorr}
\end{figure}

Fig. \ref{orcorr} and Fig. \ref{encorr} display the correlation distributions of the adjacent pixels in the R, G, and B channels for the original and encrypted pictures in diagonal, horizontal, and vertical directions, respectively.

%\begin{figure}[htbp]
%	
%	%	\vspace{-0.4cm} % ????????
%	%	\hspace{-0.2cm} % ????????
%	\centerline{\includegraphics[width=22pc]{./orencorr.png}}
%	\caption{Correlation analysis of the original and encrypted images of the R, G, B channels: (\textbf{a}) horizontal direction; (\textbf{b}) vertical direction; (\textbf{c}) diagonal direction.}
%	\label{orencorr}
%	%	\vspace*{-0.5\baselineskip}
%\end{figure}

The results show strong correlations in the original image, which are significantly decreased after encryption, indicating that the proposed algorithm can effectively resist statistical attacks.

\subsection{Differential Attack Analysis}

In order to defend against a differential attack, a secure encryption method must demonstrate susceptibility to minor alterations within the original image. Thus, every alteration of a single bit in the original image should cause the encryption process to produce an entirely distinct encrypted image \cite{9562561}. The computation of NPCR and UACI is determined according to the following formula:

%\begin{align}
%	\text{NPCR} &= \sum_{i=1}^{N} \sum_{j=1}^{M} \frac{D(i, j)}{M \times N} \times 100\%
%\end{align}
%\begin{align}
%	\text{UACI} &= \sum_{i=1}^{N} \sum_{j=1}^{M} \frac{|C_1(i,j) - C_2(i,j)|}{T \times M \times N} \times 100\% 
%\end{align}
%\begin{align}
%	\text{{D(i, j)}} &=
%	\begin{cases}
	%		0, & \text{if } C_1(i, j) = C_2(i, j) \\
	%		1, & \text{if } C_1(i, j) \neq C_2(i, j)
	%	\end{cases}
%\end{align}

\begin{figure}[htbp]
	%	\hspace{0.6cm} % 负值表示向左移动
	\centerline{\includegraphics[width=16pc]{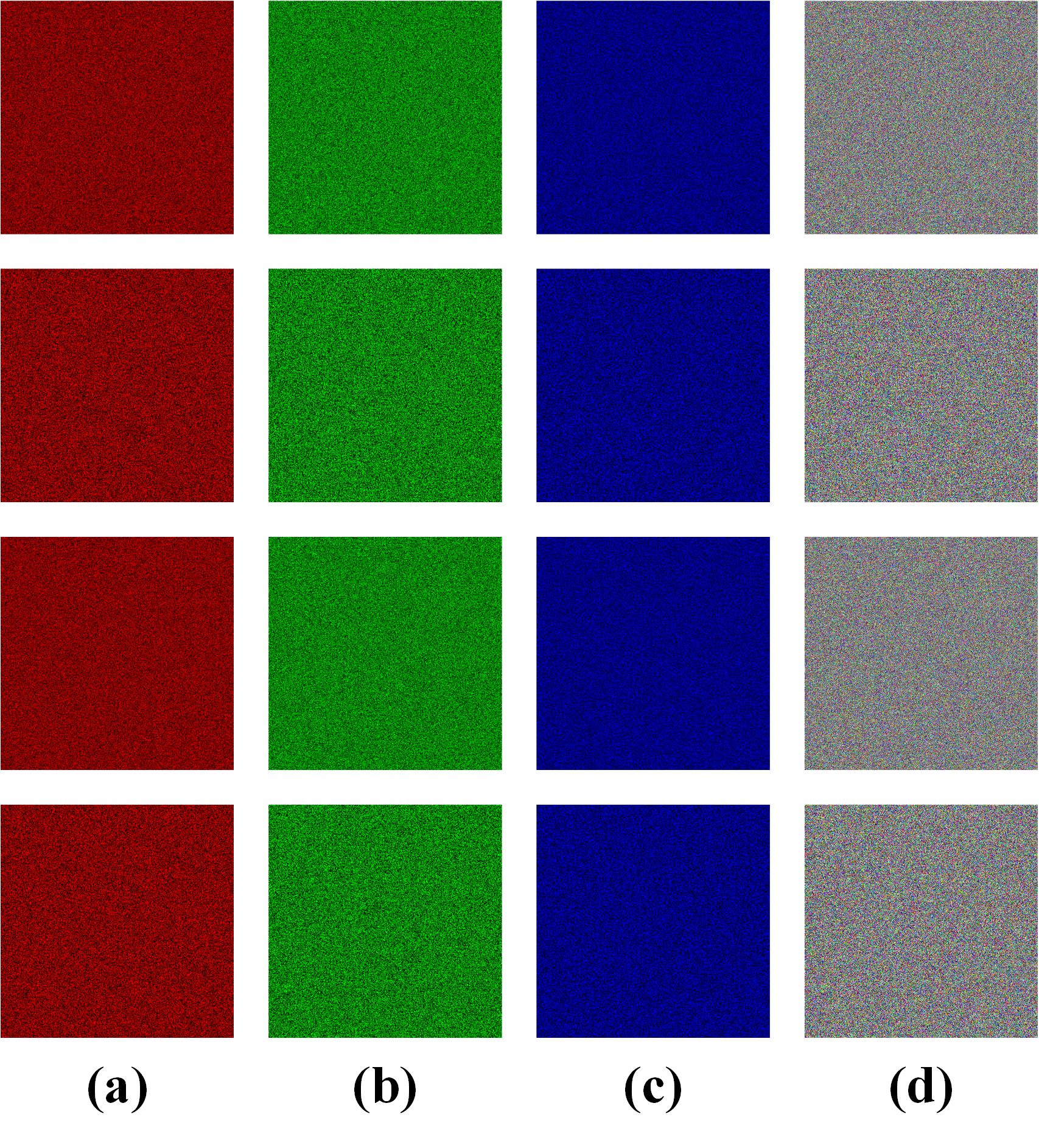}}
	\caption{Differential attack analysis: (\textbf{a}) R channel, (\textbf{b}) G channel, (\textbf{c}) B channel, (\textbf{d}) combined images.}
	\label{dif}
\end{figure}

%\begin{equation}
%	\begin{aligned}
	%		\left\{
	%		\begin{aligned}
		%			\delta(n, m) &=
		%			\begin{cases}
			%				1, & \text{if } C_1(i, j) \neq C_2(i, j) \\
			%				0, & \text{otherwise}
			%			\end{cases} \\
		%			\text{NPCR} &= \sum_{n=1}^{N} \sum_{m=1}^{M} \frac{\delta(n, m)}{M \times N}\% \\
		%			\text{UACI} &= 100\% \times \sum_{n=1}^{N} \sum_{m=1}^{M} \frac{\left| C_1(n,m) - C_2(n,m) \right|}{T \times M \times N} 
		%		\end{aligned}
	%		\right.
	%	\end{aligned}
%\end{equation}
\begin{equation}
	\begin{aligned}
		\left\{
		\begin{aligned}
			\Delta_{nm} &= |C_1(n, m) - C_2(n, m)| \\
			\text{NPCR} &= \frac{100\%}{MN} \sum_{n=1}^{N} \sum_{m=1}^{M} \text{sign}(\Delta_{nm}) \\
			\text{UACI} &= \frac{100\%}{TMN} \sum_{n=1}^{N} \sum_{m=1}^{M} \Delta_{nm}
		\end{aligned}
		\right.
	\end{aligned}
\end{equation}

where \(C_1\) and \(C_2\) represent two encrypted images of size \(M \times N\), with $T$ denoting the maximum permissible pixel intensity. The theoretical values of NPCR and UACI are 99.6094\% and 33.4635\%, respectively. After subtracting the encrypted images, the resulting images in Fig. \ref{dif} show a substantial difference.

%\begin{figure}[htbp]
%	
%	%	\hspace{0.6cm} % ????????
%	\centerline{\includegraphics[width=20pc]{./difkey.png}}
%	\caption{Differential attack and key sensitivity analysis: (\textbf{a}) R channel; (\textbf{b}) G channel; (\textbf{c}) B channel; (\textbf{d}) combined images.}
%	\label{difkey}
%	%	\vspace*{-0.5\baselineskip}
%\end{figure}
%Tab. \ref{NPCRd} presents the NPCR values in the context of differential attack analysis, and Tab. \ref{UACId} demonstrates the UACI values within this analysis.
Tab. \ref{NPCRdd} presents the NPCR and UACI values in the context of differential attack analysis. Tab. \ref{NPCRdd2} shows the comparison results of the proposed algorithm and other algorithms. It has been observed that the NPCR and UACI values are near the theoretical values. Therefore, the proposed algorithm demonstrates sensitivity to even minor variations in input images.

%\begin{table}[htbp]
%	\centering
%	\setlength{\tabcolsep}{3pt}
%	%	\vspace{-0.4cm} % 负值表示向左移动
%	\caption{NPCR of three channels in differential attack analysis.}
%	\label{NPCRd}
%	\begin{tabular}{ccccc}
	%		\toprule
	%		\textbf{Image}&	\textbf{Size}&\multicolumn{3}{c}{\textbf{NPCR(\%)}}\\
	%		\cmidrule{3-5}
	%		& & R & G& B \\
	%		\midrule
	%		1 & $304 \times 228$ &99.4518	&99.4936&99.4878\\
	%		2 & $1280\times720$ & 99.4516	&99.4630&99.4502\\
	%		3 & $960 \times 540$ & 99.4850	&99.4632&99.4606\\
	%		4 & $640 \times 360$ & 99.4653	&99.4193&99.4674\\
	%		5 & $1280 \times 720$ &99.4714	&99.4512&99.4570\\
	%		6 & $960 \times 540$ & 99.4734	&99.4622&99.4514\\
	%		\bottomrule
	%	\end{tabular}
%\end{table}
%
%\begin{table}[htbp]
%	\centering
%	\setlength{\tabcolsep}{3pt}
%	%	\vspace{-0.4cm} % 负值表示向左移动
%	\caption{UACI of three channels in differential attack analysis.}
%	\label{UACId}
%	\begin{tabular}{ccccc}
	%		\toprule
	%		\textbf{Image}&	\textbf{Size}&\multicolumn{3}{c}{\textbf{UACI(\%)}}\\
	%		\cmidrule{3-5}
	%		& & R & G& B \\
	%		\midrule
	%		1 & $304 \times 228$ &32.0422&32.0962&32.0396\\
	%		2 & $1280\times720$ & 32.0802&32.0938&32.1077\\
	%		3 & $960 \times 540$ &32.0696&32.0393&32.0721\\
	%		4 & $640 \times 360$ &31.9462&32.3061&32.2407\\
	%		5 & $1280 \times 720$& 32.0862&32.0625&32.0130\\
	%		6 & $960 \times 540$ &32.0247&32.0538&32.0022\\
	%		\bottomrule
	%	\end{tabular}
%\end{table}

\begin{table}[htbp]
	%	\scriptsize
	\centering
	\setlength{\tabcolsep}{1pt}
	%	\vspace{-0.4cm} % ????????
	\caption{NPCR and UACI of three channels in differential attack analysis.}
	\label{NPCRdd}
	\begin{tabular}{cccccccc}
		\toprule
		\textbf{Image}&	\textbf{Size}&\multicolumn{3}{c}{\textbf{NPCR(\%)}}&\multicolumn{3}{c}{\textbf{UACI(\%)}}\\
		\cmidrule{3-5}\cmidrule{6-8}
		& & R & G& B & R & G& B\\
		\midrule
		1 & $640 \times 360$ &99.6007	&99.6007&99.6007&33.4600&33.4181&33.4215\\
		2 & $1280\times720$ & 99.6055	&99.6055&99.6055& 33.5319&33.4693&33.4361\\
		3 & $960 \times 540$ & 99.6190	&99.6190&99.6190&33.3859&33.4021&33.4368\\
		4 & $640 \times 360$ & 99.5981	&99.5981&99.5981&33.4310&33.4450&33.4540\\
		5 & $1280 \times 720$ &99.5947	&99.5947&99.5947& 33.4711&33.4814&33.4958\\
		6 & $960 \times 540$ & 99.6154	&99.6154&99.6154&33.4482&33.4350&33.4233\\
		
		\bottomrule
	\end{tabular}
	%	\vspace*{-0.5\baselineskip}
\end{table}

\begin{table}[htbp]
	%	\scriptsize
	\centering
	\setlength{\tabcolsep}{1pt}
	%	\vspace{-0.4cm} % ????????
	\caption{Comparison of NPCR and UACI in differential attack analysis.}
	\label{NPCRdd2}
	\begin{tabular}{ccc}
		\toprule
		\textbf{Algorithm}&\textbf{NPCR(\%)}&\textbf{UACI(\%)}\\
		\midrule
		Ref. \cite{rezaei2023secure} & 99.6286&33.4669 \\
		Ref. \cite{BENTOUTOU2020176}& 99.61	&33.48 \\
		Ref. \cite{BEZERRA2023113160}& 99.63&33.52 \\
		Ref. \cite{zhao2023satellite}&99.61&33.43 \\
		Ref. \cite{mao2023real} & 99.6338&33.5048 \\
		Ref. \cite{LI2022111989} & 99.6147&	33.4723\\
		Ref. \cite{SU2024121088} & 99.6089&	33.4619 \\
		Ref. \cite{UMAR2024125050} & 99.6006 & 33.4295\\
		Proposed &99.6071	&33.4654\\
		
		\bottomrule
	\end{tabular}
	%	\vspace*{-0.5\baselineskip}
\end{table}

\subsection{Key Sensitive Analysis}

\begin{figure}[htbp]
	
	%	\hspace{0.6cm} % 负值表示向左移动
	\centerline{\includegraphics[width=16pc]{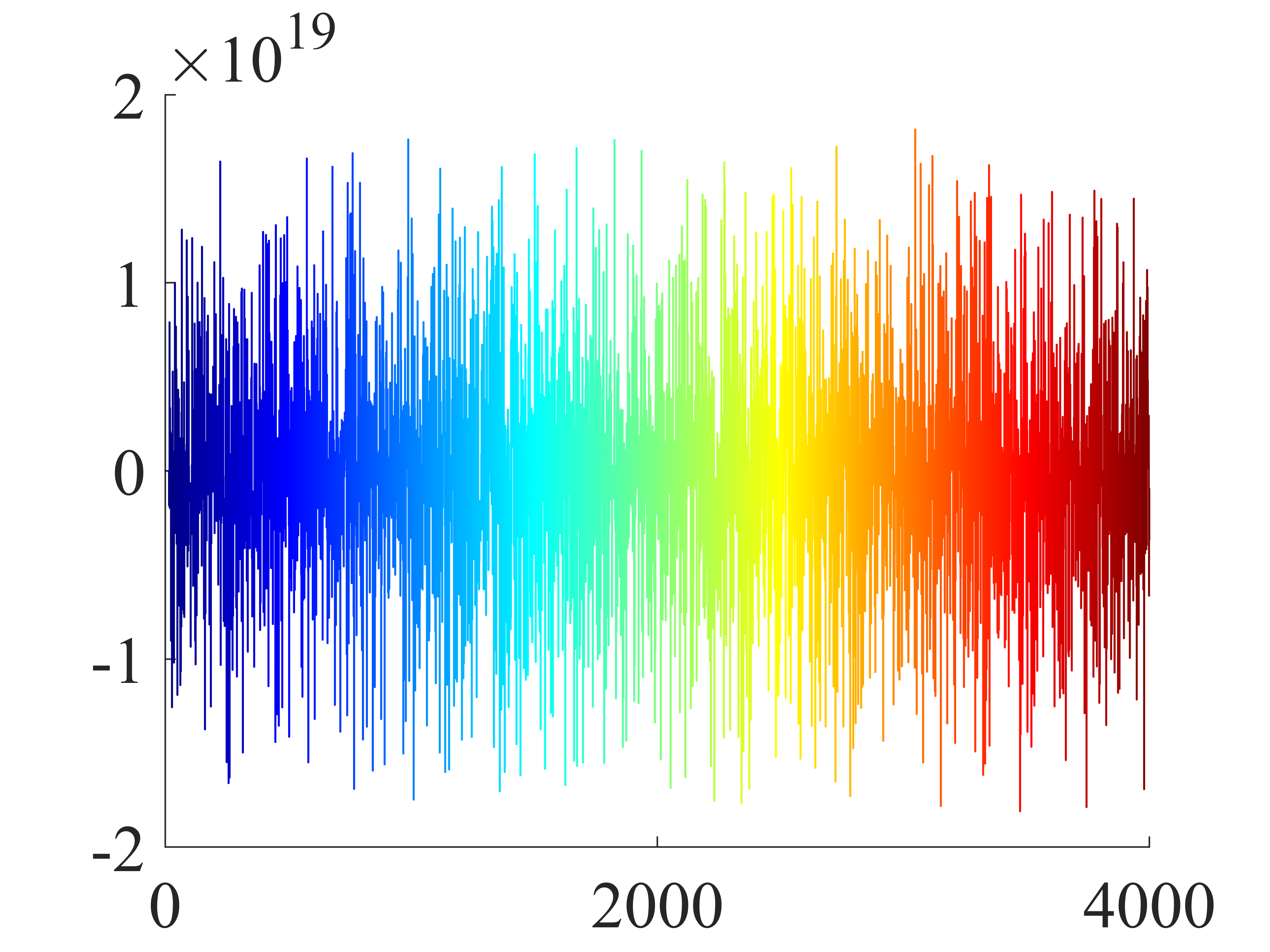}}
	\caption{The difference waveform of two keys over time.}
	\label{keyd}
\end{figure}

\begin{figure}[htbp]
	
	%	\hspace{0.6cm} % 负值表示向左移动
	\centerline{\includegraphics[width=16pc]{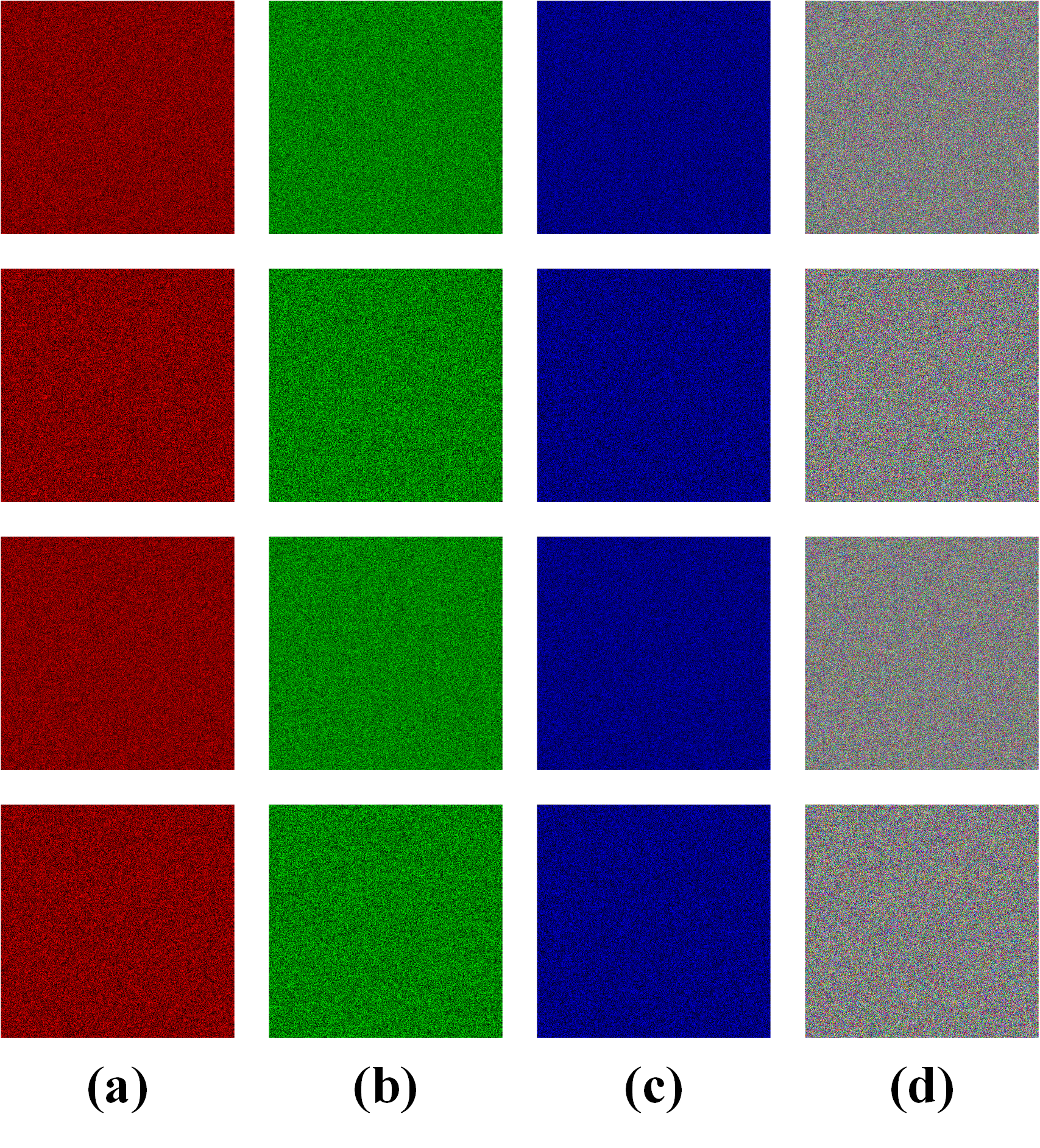}}
	\caption{Key sensitivity analysis: (\textbf{a}) R channel, (\textbf{b}) G channel, (\textbf{c}) B channel, (\textbf{d}) combined images.}
	\label{key}
\end{figure}

In order to enhance the security of an encryption algorithm against brute force assaults, it is crucial for the algorithm to demonstrate a high level of sensitivity to even the slightest modifications in the secret key \cite{tariq2020novel}. By making a little change to the original key value using just one binary bit, the key shows noticeable variations. Fig. \ref{keyd} depicts the temporal waveform of the original key and the altered key, as well as their disparities. Fig. \ref{key} displays the results of the key sensitive experiments.

%Tab. \ref{NPCRk} presents the NPCR values in the key sensitivity analysis, and Tab. \ref{UACIk} details the UACI values derived from the analysis.
Tab. \ref{NPCRkk} presents the NPCR and UACI values in the key sensitivity analysis. Tab. \ref{NPCRkk2} demonstrates the comparison results of the proposed algorithm and other algorithms. The calculated values of NPCR and UACI closely approximate their respective theoretical values, demonstrating the proposed algorithm exhibits high sensitivity to changes in the key.
%\begin{table}[htbp]
%	\centering
%	\setlength{\tabcolsep}{3pt}
%	\caption{NPCR of three channels in key sensitive analysis.}
%	\label{NPCRk}
%	\begin{tabular}{ccccc}
	%		\toprule
	%		\textbf{Image}&	\textbf{Size}&\multicolumn{3}{c}{\textbf{NPCR(\%)}}\\
	%		\cmidrule{3-5}
	%		& & R & G& B \\
	%		\midrule
	%		1 & $304 \times 228$ &99.4821&99.4431&99.3594\\
	%		2 & $1280\times720$ & 99.4546&99.4724&99.4613\\
	%		3 & $960 \times 540$ &99.4664&99.4664&99.4518\\
	%		4 & $640 \times 360$ &99.4479&99.4657&99.4280\\
	%		5 & $1280 \times 720$&99.4546&99.4724&99.4613\\
	%		6 & $960 \times 540$ &99.4664&99.4664&99.4518\\
	%		\bottomrule
	%	\end{tabular}
%\end{table}
%
%\begin{table}[htbp]
%	\centering
%	\setlength{\tabcolsep}{3pt}
%	\caption{UACI of three channels in key sensitive analysis.}
%	\label{UACIk}
%	\begin{tabular}{ccccc}
	%		\toprule
	%		\textbf{Image}&	\textbf{Size}&\multicolumn{3}{c}{\textbf{UACI(\%)}}\\
	%		\cmidrule{3-5}
	%		& & R & G& B \\
	%		\midrule
	%		1 & $304 \times 228$ &31.9260&31.9883&32.0508\\
	%		2 & $1280\times720$ & 32.0671&32.1021&32.1450\\
	%		3 & $960 \times 540$ &32.0388&32.0914&32.0199\\
	%		4 & $640 \times 360$ &31.8761&32.2281&32.2876\\
	%		5 & $1280 \times 720$&32.0492&32.0922&31.9843\\
	%		6 & $960 \times 540$ &32.0199&32.0072&32.0200\\
	%		\bottomrule
	%	\end{tabular}
%\end{table}

\begin{table}[htbp]
	%	\scriptsize
	\centering
	\setlength{\tabcolsep}{1pt}
	\caption{NPCR and UACI of three channels in key sensitive analysis.}
	\label{NPCRkk}
	\begin{tabular}{cccccccc}
		\toprule
		\textbf{Image}&	\textbf{Size}&\multicolumn{3}{c}{\textbf{NPCR(\%)}}&\multicolumn{3}{c}{\textbf{UACI(\%)}}\\
		\cmidrule{3-5}\cmidrule{6-8}
		& & R & G& B & R & G& B\\
		\midrule
		1 & $640 \times 320$ &99.6120&99.6120&99.6120&33.4879&33.4846&33.4306\\
		2 & $1280\times720$ & 99.6212&99.6212&99.6212& 33.5341&33.4856&33.4512\\
		3 & $960 \times 540$ &99.6078&99.6078&99.6078&33.4882&33.4967&33.5075\\
		4 & $640 \times 360$ &99.6085&99.6085&99.6085&33.4979&33.4818&33.4700\\
		5 & $1280 \times 720$&99.6110&99.6110&99.6110&33.4418&33.4475&33.4427\\
		6 & $960 \times 540$ &99.6035&99.6035&99.6035&33.4803&33.4763&33.4969\\

		\bottomrule
	\end{tabular}
	%	\vspace*{-0.5\baselineskip}
\end{table}

\begin{table}[htbp]
	%	\scriptsize
	\centering
	\setlength{\tabcolsep}{1pt}
	%	\vspace{-0.4cm} % ????????
	\caption{Comparison of NPCR and UACI in key sensitive analysis.}
	\label{NPCRkk2}
	\begin{tabular}{ccc}
		\toprule
		\textbf{Algorithm}&\textbf{NPCR(\%)}&\textbf{UACI(\%)}\\
		\midrule
		Ref. \cite{BEZERRA2023113160}& 99.60&33.44 \\
		Ref. \cite{UMAR2024125050} & 99.5999 & 33.5060\\
		Ref. \cite{SHENG2023103592} & 99.612&	33.468\\
		Ref. \cite{JASRA2022117861} & 99.61	&33.78 \\
		Proposed &99.6052	&33.4654\\
		
		\bottomrule
	\end{tabular}
	%	\vspace*{-0.5\baselineskip}
\end{table}

%The results in Tab. \ref{NPCRkk} demonstrate that the calculated values of NPCR and UACI closely approximate their respective theoretical values. It is evident that the proposed algorithm exhibits high sensitivity to changes in the key.

%\subsection{Cropping attack analysis}
%
%Attackers frequently employ techniques like cropping and noise attacks to compromise the integrity of transmitted information, hindering the receiver from successfully decrypting the ciphertext and obtaining accurate decryption information\cite{XIAN2020106202}. The experiment focused on countering cropping assaults. Fig. \ref{crop} shows the decryption results of the cropped encrypted images at sizes 1/16, 1/8, and 1/4. These results demonstrate the effectiveness of the proposed approach in successfully resisting cropping attacks.
%
%%\begin{figure}[H]
%%	
%%	%	\hspace{0.6cm} % 负值表示向左移动
%%	\centerline{\includegraphics[width=20pc]{./crop1.png}}
%%	\caption{Result of cropping attack analysis.}
%%	\label{crop1}
%%\end{figure}
%%
%%\begin{figure}[H]
%%	
%%	%	\hspace{0.6cm} % 负值表示向左移动
%%	\centerline{\includegraphics[width=20pc]{./crop2.png}}
%%	\caption{Result of cropping attack analysis.}
%%	\label{crop2}
%%\end{figure}
%
%\begin{figure}[htbp]
%	
%	\centerline{\includegraphics[width=22pc]{./crop.png}}
%	\caption{Result of cropping attack analysis: (\textbf{a},\textbf{b},\textbf{c}) cropped encrypted images; (\textbf{d},\textbf{e},\textbf{f}) decrypted images.}
%	\label{crop}
%\end{figure}
\subsection{Ablation Analysis}

In the ablation experiments, we tested the proposed scheme using only a single chaotic sequence, referred to as the Only Chaotic sequence (\ref{chaos}) (OC\ref{chaos}) and the Only Chaotic sequence (\ref{chaos2}) (OC\ref{chaos2}). Additionally, since the generated sequences are identical across the R, G, and B channels, we also tested a scheme that generates a pseudo-random sequence of length $\frac{3 \times m \times n}{8}$, referred to as 3L. Tab. \ref{abla} shows the ablation analysis result.

\begin{table}[htbp]
	%	\scriptsize
	\centering
	\setlength{\tabcolsep}{1pt}
	%	\vspace{-0.4cm} % ????????
	\caption{Results of ablation analysis.}
	\label{abla}
	\begin{tabular}{cccccc}
		\toprule
		\textbf{Algorithm}&\textbf{NPCR(\%)}&\textbf{UACI(\%)}&\multicolumn{3}{c}{\textbf{Correlation}}\\
		\cmidrule{4-6}
		&&& Horiz. & Vert. & Diag.\\
		\midrule
		OC\ref{chaos} & 99.4781&32.0991&0.0015 &  0.0437 &  0.0013\\
		OC\ref{chaos2}&99.4476&32.1821&\textbf{-0.0006} &  0.0439 &  0.0029 \\
		3L&99.5999&33.4706&-0.0028 &  0.0052 & \textbf{-0.0001} \\
		Proposed &\textbf{99.6071}	&\textbf{33.4654}&-0.0028 &  \textbf{0.0039} & -0.0013\\
		\bottomrule
	\end{tabular}
	%	\vspace*{-0.5\baselineskip}
\end{table}

It can be observed that when only a single chaotic sequence is used, the encrypted images exhibit higher vertical correlation, and the NPCR and UACI values are less ideal compared to the proposed scheme. Furthermore, using a pseudo-random sequence three times longer does not result in significantly improved encryption performance. Considering both efficiency and security, we chose the proposed encryption scheme.

\subsection{NIST Test}

We have chosen the NIST test suite to assess the unpredictability of the data. The analysis consists of 15 tests to fulfill crucial requirements \cite{gafsi2020improved}. In order to pass the test, the p-value must exceed 0.01. The test result of an encrypted image is shown in Tab. \ref{NIST}, demonstrating the effectiveness of the proposed encryption method.

\begin{table}[htbp]
	\centering
	\setlength{\tabcolsep}{3pt}
	\caption{Result of NIST tests~suite.}
	\label{NIST}
	\begin{tabular}{ccc}
		\toprule
		\textbf{Test} &  \textbf{\emph{p}-Value}& \textbf{Assessment}\\
		\midrule
		ApproximateEntropy&	0.213309&	SUCCESS\\
		BlockFrequency&	0.350485&	SUCCESS\\
		CumulativeSums&	0.739918&	SUCCESS\\
		FFT	&0.066882&	SUCCESS\\
		Frequency&	0.213309&	SUCCESS\\
		LinearComplexity&	0.213309&	SUCCESS\\
		LongestRun&	0.350485&	SUCCESS\\
		NonOverlappingTemplate&	0.999438&	SUCCESS\\
		OverlappingTemplate&	0.739918&	SUCCESS\\
		RandomExcursions&	0.637119&	SUCCESS\\
		RandomExcursionsVariant&	0.834308&	SUCCESS\\
		Rank&	0.834308&	SUCCESS\\
		Runs&	0.911413&	SUCCESS\\
		Serial&	0.991468&	SUCCESS\\
		Universal&	0.122325&	SUCCESS\\
		\bottomrule
	\end{tabular}
\end{table}

\subsection{DIEHARD test}
Tab. \ref{die2} reveals the DIEHARD test result of an encrypted image, indicating that the encrypted image exhibits highly random behavior.
%
%\begin{table}[htbp]
%	\tiny
%	\centering
%	\setlength{\tabcolsep}{1pt}
%	\caption{Result of NIST and DIEHARD tests~suite.}
%	\label{N}
%	\begin{tabular}{ccccc}
	%		\toprule
	%		\multicolumn{2}{c}{\textbf{NIST}}&\multicolumn{2}{c}{\textbf{DIEHARD}}&Assessment\\
	%		\cmidrule{1-2}\cmidrule{3-4}
	%		Test &  \emph{p}-Value&Test &  \emph{p}-Value&\\
	%		\midrule
	%		ApproximateEntropy&	0.2133&Birthday spacing &0.3533 & PASSED \\
	%		BlockFrequency&	0.3505&Overlapping permutation & 0.7423 & PASSED \\
	%		CumulativeSums&	0.7399&Binary rank 32{$\times$}32 & 0.3813& PASSED \\
	%		FFT	&0.0669&	Binary rank 6{$\times$}8 & 0.2534& PASSED \\
	%		Frequency&	0.2133&	Bitstream & 0.8643 & PASSED \\
	%		LinearComplexity&	0.2133&	OPSO &  0.9603& PASSED \\
	%		LongestRun&	0.3505&	OQSO &  0.0897& PASSED \\
	%		NonOverlappingTemplate&	0.9994&	DNA &  0.9930 & PASSED \\
	%		OverlappingTemplate&	0.7399&	Count the ones 01&0.3173 & PASSED \\
	%		RandomExcursions&	0.6371&	Count the ones 02&0.4699& PASSED \\
	%		RandomExcursionsVariant&	0.8343&	Parking lot&0.9637& PASSED \\
	%		Rank&	0.8343&	2DS sphere&0.4767& PASSED \\
	%		Runs&	0.9114&	3DS spheres&0.5744 & PASSED \\
	%		Serial&	0.9915&	Squeeze&0.2783 & PASSED \\
	%		Universal&	0.1223&	Overlapping sum&0.1785& PASSED \\
	%		&&Runs&0.7988& PASSED \\
	%		&&Craps&0.6144 & PASSED\\
	%		
	%		\bottomrule
	%	\end{tabular}
%	%	\vspace*{-0.5\baselineskip}
%\end{table}

\begin{table}[htbp]
	\centering
	\setlength{\tabcolsep}{3pt}
	\caption{Result of DIEHARD tests~suite.}
	\label{die2} 
	\begin{tabular}{ccc}
		\toprule
		\textbf{Test} &  \textbf{\emph{p}-Value}& \textbf{Assessment}\\
		\midrule
		Birthday spacing &0.3533 & PASSED \\
		
		Overlapping permutation & 0.7423 & PASSED \\
		
		Binary rank 32{$\times$}32 & 0.3813& PASSED  \\ 
		
		Binary rank 6{$\times$}8 & 0.2534& PASSED  \\
		
		Bitstream & 0.8643 & PASSED \\
		
		OPSO &  0.9603& PASSED \\
		
		OQSO &  0.0897& PASSED\\
		
		DNA &  0.9930 & PASSED\\
		
		Count the ones 01&0.3173 & PASSED\\
		
		Count the ones 02&0.4699& PASSED\\
		
		Parking lot&0.9637& PASSED\\
		
		2DS sphere&0.4767& PASSED \\
		
		3DS spheres&0.5744 & PASSED\\
		
		Squeeze&0.2783 & PASSED\\
		
		Overlapping sum&0.1785& PASSED \\
		
		Runs&0.7988& PASSED \\
		
		Craps&0.6144 & PASSED\\
		\bottomrule
	\end{tabular}
\end{table}

\section{Conclusion}\label{sec6}

Implementing real-time video encryption on satellite embedded systems is essential for improving the security and dependability of satellite communications. Utilizing existing video encryption systems on satellite payloads presents obstacles such as real-time performance and limited computing resources. In contrast, this work presents a novel approach for encrypting videos based on 1D chaotic mapping, specifically designed to address these limitations. We deployed a real-time video encryption algorithm on a satellite for the first time, demonstrating exceptional performance. According to the results of the experiments, this approach exhibits robust real-time performance and requires low power consumption. Notably, while considering only the encryption scheme, the system is capable of processing 2K video in real time, demonstrating its applicability to high-resolution video scenarios. Furthermore, FPGA deployment verified that the chaotic mapping values are consistent with those obtained on a Raspberry Pi, confirming the method’s hardware feasibility and stability. Comprehensive security evaluations also confirmed the system’s capability to endure diverse attacks. In the future, we will explore the extension of this method to support higher-resolution videos and integrate the encryption scheme with real satellite applications to further validate its effectiveness and adaptability. It is expected that this study will advance the field of video streaming security for satellite embedded devices and lay a strong foundation for future advancements in secure, real-time data transmission in satellite communication systems.

\section*{ACKNOWLEDGMENT}

This study is supported by the Talent Scientific Fund of Lanzhou University. Experiments for this study were facilitated by the TianSuan Constellation Platform at Beijing University of Posts and Telecommunications.

Special thanks are extended to Professor Shangguang Wang at Beijing University of Posts and Telecommunications for his invaluable guidance and support. Appreciation is also due to PhD student Ruolin Xing and Master's student Linshuo Song at Beijing University of Posts and Telecommunications for their significant contributions to the research.
\bibliography{refs}
\textbf{\bibliographystyle{IEEEtran}}

\end{document}